\documentclass[%
 reprint,
 floatfix,
superscriptaddress,
longbibliography,
 amsmath,amssymb,
 aps,
 prx,
]{revtex4-2}
\makeatletter
\def\set@curr@file#1{%
 \begingroup
 \escapechar\m@ne
 \xdef\@curr@file{\expandafter\string\csname #1\endcsname}%
 \endgroup
}
\def\quote@name#1{"\quote@@name#1\@gobble""}
\def\quote@@name#1"{#1\quote@@name}
\def\unquote@name#1{\quote@@name#1\@gobble"}
\makeatother
\usepackage{graphicx}
\graphicspath{ {./images/} }
\usepackage{dcolumn}
\usepackage{bm}
\usepackage{xfrac}
\usepackage{siunitx}
\usepackage{amsmath}
\usepackage[utf8]{inputenc}
\usepackage[T1]{fontenc}
\usepackage[export]{adjustbox}
\usepackage[colorinlistoftodos,prependcaption,textsize=tiny]{todonotes}
\usepackage{gensymb}
\presetkeys%
 {todonotes}%
 {inline, backgroundcolor=yellow}{}

\begin{document}
\preprint{APS/123-QED}

\title{Single-Bubble Dynamics in Nanopores:\\Transition Between Homogeneous and Heterogeneous Nucleation}

\author{Soumyadeep Paul}
\affiliation{%
Department of Mechanical Engineering, The University of Tokyo 7-3-1, Hongo, Bunkyo-ku, Tokyo 113-8656, Japan
}%
\author{Wei-Lun Hsu}%
\affiliation{%
Department of Mechanical Engineering, The University of Tokyo 7-3-1, Hongo, Bunkyo-ku, Tokyo 113-8656, Japan
}%
\author{Mirco Magnini}%
\affiliation{Department of Mechanical Engineering, University of Nottingham, Nottingham, NG7 2RD, United Kingdom
}%
\author{Lachlan R. Mason}
\affiliation{Data-Centric Engineering Programme, The Alan Turing Institute, London, NW1 2DB, United Kingdom}
\author{Ya-Lun Ho}%
\affiliation{%
Department of Mechanical Engineering, The University of Tokyo 7-3-1, Hongo, Bunkyo-ku, Tokyo 113-8656, Japan
}%
\author{Omar K. Matar}%
\affiliation{%
Department of Chemical Engineering, Imperial College London, London, SW7 2AZ, United Kingdom
}%
\author{Hirofumi Daiguji}%
 \email{Corresponding author: daiguji@thml.t.u-tokyo.ac.jp}
\affiliation{%
Department of Mechanical Engineering, The University of Tokyo 7-3-1, Hongo, Bunkyo-ku, Tokyo 113-8656, Japan
}%





\begin{abstract}
When applying a voltage bias across a thin nanopore, localized Joule heating can lead to single-bubble nucleation, offering a unique platform for studying nanoscale bubble behavior, which is still poorly understood. Accordingly, we investigate bubble nucleation and collapse inside solid-state nanopores filled with electrolyte solutions and find that there exists a clear correlation between homo/heterogeneous bubble nucleation and the pore diameter.
As the pore diameter is increased from 280~nm to 525~nm, the nucleation regime transitions from predominantly periodic homogeneous nucleation to a non-periodic mixture of homogeneous and heterogeneous nucleation. A transition barrier between the homogeneous and heterogeneous nucleation regimes is defined by considering the relative free-energy costs of cluster formation. A thermodynamic model considering the transition barrier and contact-line pinning on curved surfaces is constructed, which determines the possibility of heterogeneous nucleation. It is shown that the experimental bubble generation behavior is closely captured by our thermodynamic analysis, providing important information for controlling the periodic homogeneous nucleation of bubbles in nanopores.

\begin{description}
\item[Keywords]
Ionic Joule heating, Solid-state nanopores, Periodic bubble generation, Seed bubbles,\\ Microchannel heat sink, Hetero/homogeneous bubble nucleation, Cluster ripening
\end{description}
\end{abstract}

\maketitle


\section{\label{sec:level1}INTRODUCTION}



Following Moore's Law, transistors continue to shrink, enabling the miniaturization of electronic chips. In the meantime, these closely packed electronic components are subject to significant heat generation in high-performance electronic devices. In this context, two-phase cooling using microchannel evaporators has emerged as an energy-efficient cooling solution. In this system, liquid refrigerant enters from one end of a channel and vaporizes while consuming heat from the underlying chip. However, this heat transfer method via flow boiling may cause chip overheating due to flow instabilities originating from spontaneous bubble nucleation on the channel walls. Although previous studies have shown that boiling incipience and its associated instabilities could be effectively eliminated through wall nucleation suppression and microheater bubble seeding~\cite{thome2007bubble,Liu2010,Xu2009}, controlling the unpinning and departure dynamics of heterogeneous nucleating bubbles from the heater surfaces remains challenging. For compact systems (e.g.,~3D chips) comprising sub-100-µm channels, homogeneous bubble seeds, originating in the bulk phase and operating on the nanoscale, are needed.

Homogeneous nanobubble~\cite{Nagashima2014,Levine2016} generation inside nanopores can serve as a potential method for bubble seeding. Through the application of a high-voltage bias across a solid-state nanopore immersed in a concentrated salt solution, intense and localized Joule heat is produced inside the nanopore. After the initial heating, the solution inside the nanopore becomes metastable, following which a bubble nucleates homogeneously under a superheating condition.

Golovchenko and coworkers~\cite{Nagashima2014,Levine2016} showed that homogeneous nucleation inside the nanopore leads to periodic and uniformly sized nanobubbles, making it attractive for generation of a well-organized succession of bubbly, slug, and annular flow regimes in microchannels.
Not only do these homogeneous nanopore bubbles hold tremendous potential for electronic cooling~\cite{Liu2010}, but they could also play a role in various applications ranging from biomedical imaging~\cite{Pellow2017c} to froth flotation~\cite{FAN2010}.
However, depending on conditions, heterogeneous bubble nucleation, which requires a lower nucleation energy~\cite{Witharana2012, Wang2018}, can simultaneously occur on the inner walls of the nanopore, reducing the efficiency of homogeneous bubble emission.
Despite the fact that bubble nucleation on metallic nanoparticles~\cite{Wang2018}, nanorods~\cite{boulais2013plasma}, and nanodots~\cite{Carlson2012} has gained significant attention in recent years due to its potential game-changing applications in photothermal cancer therapy~\cite{lukianova2014demand} and solar thermal powerplants~\cite{baral2014comparison,neumann2012solar}, the exact mechanisms of homo/heterogeneous nucleation and pinning or confinement effects are still under debate, and an in-depth understanding is lacking.

\begin{figure}
\includegraphics[height=3.9785 in,width=3.136 in,angle=0]{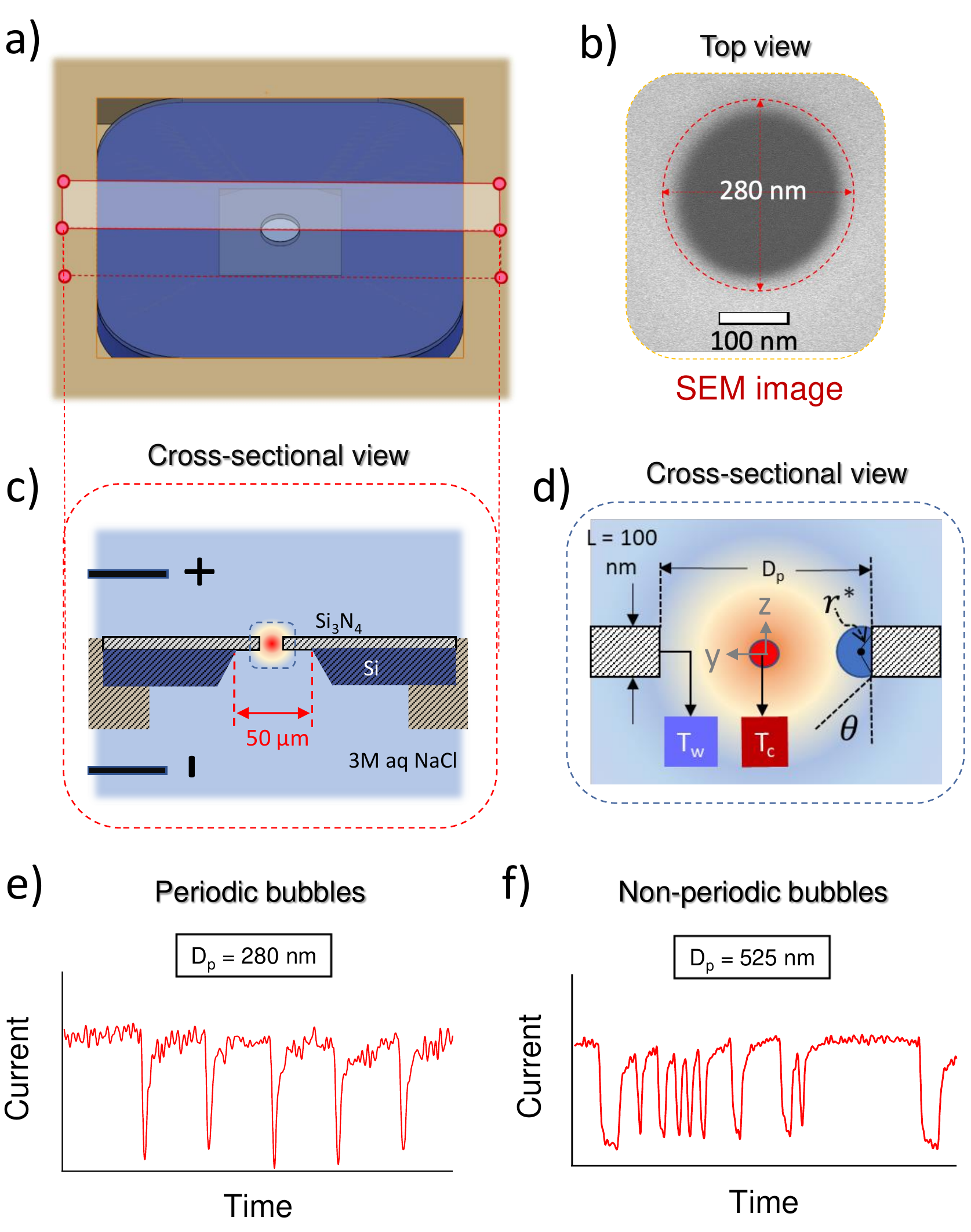}
\caption{\label{fig:1} (a) 3D image of a nanopore (a cylindrical hole on a silicon nitride thin layer) where the silicon chip is affixed on a fluidic tank. (b) SEM image of the $D_\text{p}=280$ nm nanopore. (c) Schematic (in a cross-sectional view) of the experimental setup. Bias voltages applied across the nanopore, resulting in localized Joule heating inside the nanopore. (d) Close-up view of the nanopore in (c) showing homogeneous and heterogeneous clusters formation at the pore center and cylindrical pore surface respectively. (e) Uniform quasi-periodic signals were observed when there was only homogeneous nucleation. (f) Periodicity was lost when heterogeneous nucleation started on the cylindrical pore surface.}
\end{figure}

In this paper, we investigate the bubble generation mechanism in nanopores due to Joule heating effects to control the occurrence of homogeneous and heterogeneous bubble nucleation.
We study bubble generation in 280-nm and 525-nm nanopores (Fig.~\ref{fig:1}(e,f)) and show that in the smaller pores, homogeneous nucleation is dominant, giving rise to periodic bubbles, while the bigger pores yield a combination of homogeneous and heterogeneous bubbles, eliminating the periodicity. We introduce a thermodynamic parameter, the transition barrier ($\xi$), to elucidate the transition dynamics between homogeneous and heterogeneous bubble nucleation. In addition to $\xi$, a second parameter of interest, $\zeta$, is established that captures the effect of nanoconfinement due to limited available surface area for heterogeneous nucleation on the cylindrical pore surface. Both these parameters influence the contact angle and the nucleation temperature of the heterogeneous bubbles. This analysis will shed light on the transition mechanism between homogeneous and heterogeneous nucleation. 

\section{EXPERIMENTAL METHODS}

Circular nanopores (as seen in the scanning electron microscope (SEM) image shown in Fig.~\ref{fig:1}(b) and Supplemental Material Fig.~S1.1~\cite{supp}) were made on silicon nitride chips (Model No.~4088SN-BA) purchased from Alliance Biosystems Inc., each comprising a 100-nm-thick silicon nitride (Si$_{3}$N$_{4}$) membrane deposited on a 200-µm-thick silicon substrate with an approximately square $\SI{50}{\micro\meter} \times \SI{50}{\micro\meter}$ opening at the center. The nanopores were etched at the center of the free-standing part of the membrane using a focused ion beam (SMI3050: SII Nanotechnology, a precisely focused $``$fine$ "$ Ga$^{+}$ beam at around 10--17~pA). Before performing the experiments, the nanopore-containing chip was boiled in a piranha solution to remove organic contaminants. The chip was then glued between two additively manufactured fluidic tanks such that the nanopore became the only fluidic connection [Fig.~\ref{fig:1}(a,c)], and these fluidic tanks were filled with 50\% ethanol solution to ensure that there were no air gaps in the nanopore system. A 3M NaCl electrolyte solution was prepared by diluting 5M NaCl (Sigma-Aldrich) with deionized (DI) water. Before transferring the electrolyte into the fluidic tanks, they were flushed with DI water multiple times, followed by flushing with 3M NaCl multiple times to remove any traces of ethanol or water in the system. After these steps, the fluidic tanks were filled with the prepared 3M NaCl solution. Ag/AgCl electrodes were inserted into the tanks, and square voltage pulses were applied across the nanopore by means of a function generator (Tektronix AFG3151C). The circuit current was determined by measuring the voltage across a shunt resistor using a high-bandwidth oscilloscope (Tektronix MDO3052). Downward current spikes were observed when bubbles were generated, blocking ion transport through the nanopore. Analysis of the nucleation time and bubble duration based on current variations revealed the nucleation conditions and bubble behavior in the nanopore. \par

During the experiments, collapse of bubbles in the vicinity of the nanopore can lead to shock or stress wave formations and nanojets, which may erode the surface and thus alter the roughness of the pore surface. For example, for the 280 nm diameter pore, we found that the baseline current increased after prolonged bubble generation experiments, implying an expansion of the pore diameter due to cavitation erosion. However, homogeneous bubble nucleation scheme still maintained under specific voltage ranges, highlighting that the surface roughness on the pore surface may not be the primary factor behind homo/heterogeneous transition (Supplemental Material, Sec.~7~\cite{supp}). Also, through simulations we considered nanoscale defects on the pore surface and found their effects on the temperature variation on pore surface are marginal. As the transition between homogeneous and heterogeneous nucleation is primarily governed by the nanopore temperature distribution, surface roughness effects can be neglected in the current system. \par

\section{THEORETICAL ANALYSIS}
Due to the high magnitude of the bias voltages and the small thickness of the dielectric membrane, an intense electric field was generated in the nanopore, which was converted into remarkable Joule heat ($H$) following $H = \textbf{J}\cdot\textbf{E}$. Here, $\textbf{J}=\sigma\textbf{E}$ is the current density when the advection current is negligible; $\textbf{E}$ is the local electric field, and $\sigma$ is the ionic conductivity. This Joule heat is transferred to the surrounding electrolyte as well as to the silicon nitride membrane, resulting in a sharp thermal gradient in the vicinity of the membrane wall in the solution. As the nanopore Joule heating progresses, the pore center temperature, $T_\text{c}$ rises at a higher rate than the wall temperature, $T_\text{w}$, leading to $T_\text{c} > T_\text{w}$ (Fig.~\ref{fig:1}d). Prior to the Joule-heating process, $T_\text{c} = T_\text{w}= T_\text{0}$, where $T_\text{0}$ is the ambient temperature.\par

Unlike previous studies on microheater bubbles~\cite{Nguyen2018, li2008dynamic, yin2004bubble} where the bulk temperature is always lower than the surface temperature, thereby allowing only heterogeneous nucleation, in nanopore Joule heating, a higher liquid temperature at the pore center also allows the possibility of homogeneous nucleation. For homogeneous nucleation, a higher superheating temperature is required to satisfy the kinetic requirement for cluster formation ($T_\text{c}>$~575~K for water~\cite{Avedisian1985}), whereas heterogeneous nucleation temperatures largely depend upon the dimensions of the nanostructures on the nucleating surface, following the Young--Laplace equation~\cite{Witharana2012} (424~K to 474~K for decreasing nanostructure diameters from 500~nm to 100~nm). \par

The cross-pore radial temperature difference, expressed as $\Delta T_\text{p} = T_\text{c} - T_\text{w}$, dictates whether nanopore Joule heating will result in a homogeneous or heterogeneous bubble. $\Delta T_\text{p}$ can be controlled by varying the pore diameter and bias voltage. To obtain the temperature distribution responsible for the bubble behavior, numerical simulations were employed to solve the energy-conservation equation

\begin{eqnarray}
\rho c_\text{p}\frac{\partial T}{\partial t} &&= \frac{1}{y}\frac{\partial}{\partial y}\left(\kappa y \frac{\partial T}{\partial y}\right) +\frac{\partial}{\partial z}\left(\kappa \frac{{\partial}T}{{\partial} z}\right) + \sigma{\mid\textbf{E}\mid}^2,
\label{eq:one}
\end{eqnarray}
where $\rho$, $c_\text{p}$, and $\kappa$ denote the temperature-dependent water density, specific heat, and thermal conductivity, respectively, and $t$ represents time. On an axisymmetric reference frame centered at the center of the nanopore, $y$ represents the radial coordinate and $z$ the axial coordinate. Equation~(\ref{eq:one}) was solved on a finite-volume mesh with appropriate boundary conditions and numerical discretizations (Supplemental Material, Sec.~2~\cite{supp}).\par
\begin{figure}
	\includegraphics[height=2.72 in,width=3.4 in,angle=0]{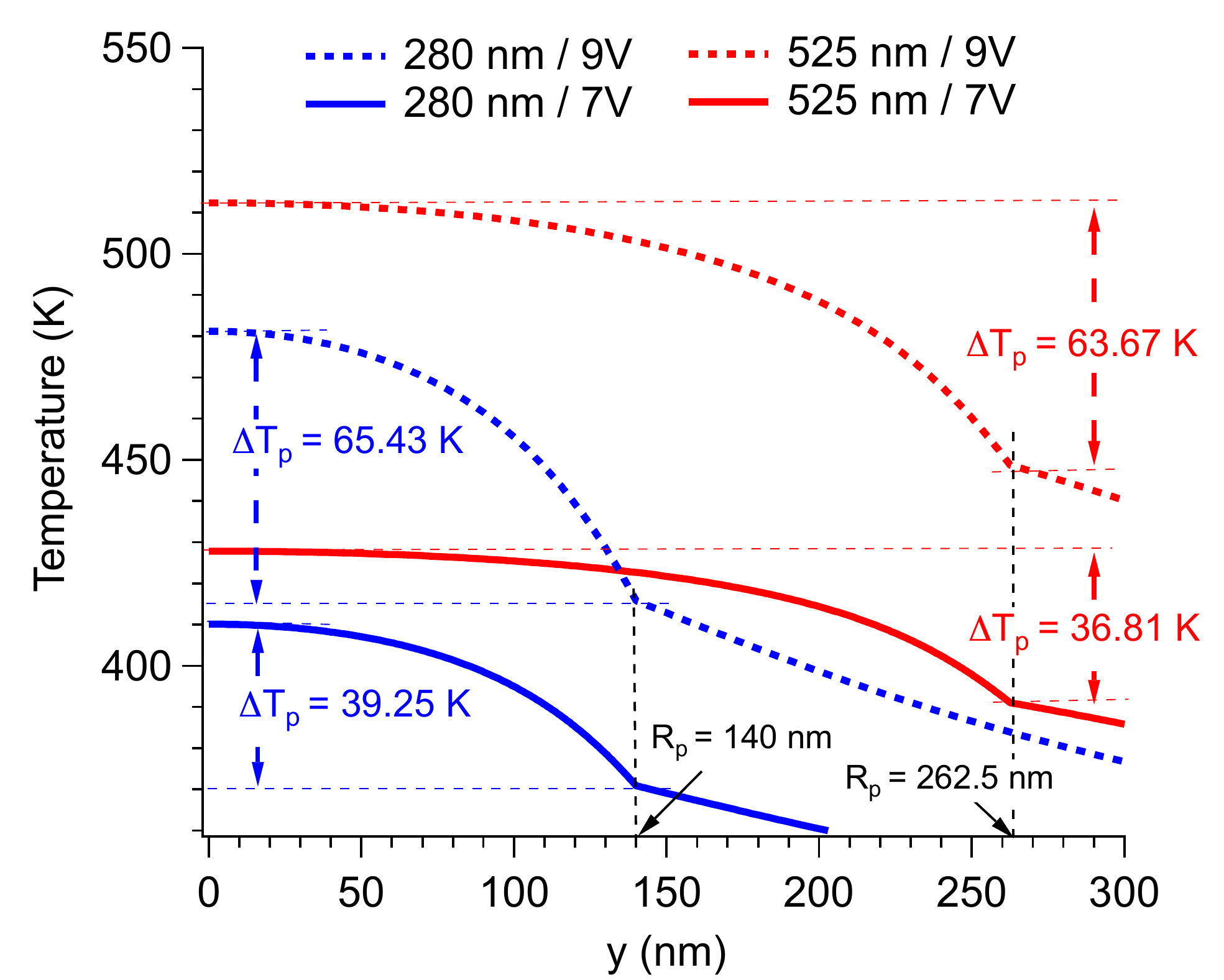}
	\caption{\label{fig:2} Temperature distributions across pore cross-section for $D_\text{p} =$ 280~nm and 525~nm after $\SI{20}{\micro\s}$ of Joule heating under bias voltages of 7~V and 9~V, with $T_\text{0}$ = 300.15~K. $y$ denotes the radial position from the pore centerline, as shown in Fig.~\ref{fig:1}(d). $R_\text{p} = D_\text{p}/2$ is the pore radius. The value of $\sigma$ in Eq.~(\ref{eq:one}) was defined at $T_\text{0}$, which remained constant during the simulation.}
\end{figure}
Figure~\ref{fig:2} shows the simulated nanopore cross-sectional temperature distributions for pores of $D_\text{p} =$ 280~nm and 525~nm obtained after $\SI{20}{\micro\s}$ of Joule heating for bias voltages of 7~V and 9~V. We can draw two major observations: i) a higher voltage leads to a higher temperature rise and higher $\Delta T_\text{p}$; and ii) for the same voltage, a smaller pore results in a higher value of $\Delta T_\text{p}$ ($\Delta T_\text{p280} - \Delta T_\text{p525}$ is 2.44~K and 1.76~K for $V_\text{app} =$ 7~V and $V_\text{app} =$ 9~V, respectively). We also simulated the effect of varying membrane thickness (pore length) (Fig.~\ref{fig:1}d) such that $L=\{50,75,100\}$ nm while $D_\text{p}$ and $V_\text{app}$ are kept constant at 280 nm and 7~V respectively. We concluded that decreasing $L$ results in a more rapid temperature rise but $\Delta T_\text{p}$ reduces for a given $T_\text{w}$. This is due to the excess heating near the edge of the cylindrical pore, where the electric field is higher. This edge effect increases with increasing the $\sfrac{D_\text{p}}{L}$ ratio.\par
It should be noted that, for simplicity, we neglected the effect of temperature on the electrical conductivity in this section. In the complete Joule-heating model, as explained in the Supplemental Material, Sec.~1~\cite{supp}, $\sigma$ increases with $T$, following an empirical relationship established in a previous study~\cite{Levine2016}. For our pore and voltage configurations, we tuned this $\sigma$--$T$ relationship to fit the experimentally observed nanopore current in our simulations (Supplemental Material, Secs.~2 and 3~\cite{supp}). In future sections, we incorporate this relationship into the Joule-heating model to capture the nanopore temperature distributions. As $T_\text{c} > T_\text{w}$, when the $\sigma$--$T$ variation is considered, Joule heat intensity at the pore center would increase by a higher amount than that near the cylindrical pore surface, resulting in an even higher $\Delta T_\text{p}$. The transient variations of $T_\text{c}$ and $T_\text{w}$ for the pore configurations under study are shown in the Supplemental Material, Sec.~4~\cite{supp}.\par
The average electric field magnitude inside the nanopore can be approximately obtained from the nanopore circuit model~\cite{Gadaleta2014} as
\begin{eqnarray}
E_\text{p} = \frac{V_\text{app}}{(L+\frac{\pi D_\text{p}}{4})},
\label{eq:two}
\end{eqnarray}
Equation~(\ref{eq:two}) shows that when increasing the bias voltage, $V_\text{app}$, and decreasing the pore diameter, $D_\text{p}$, the electric field within the nanopore intensifies, causing more intense Joule heating. If we consider a uniform electric field within the nanopore and neglect the temperature dependency of the thermophysical quantities and diffusion in the $z$ direction, the radial temperature difference across the nanopore, $\Delta T_\text{p}$, can be expressed as
\begin{equation}
\Delta T_\text{p}(t) = D_\text{T}\int_0^t \left[{{\nabla}}^2(T_\text{c} - T_\text{w})\right]dt,
\label{eq:three}
\end{equation}
where ${{\nabla}}^2 = \frac{1}{y}\frac{\partial}{\partial y}\left(y\frac{\partial }{\partial y}\right)$ is the $y$ component of the Laplacian operator, $T_\text{c}$ and $T_\text{w}$ denote the nanopore temperatures near the pore center and cylindrical pore surface, respectively, and $D_\text{T} = \sfrac{\kappa}{\rho c_\text{p}}$ denotes the thermal diffusivity. In a steady state, ${{\nabla}}^2T_\text{c} = {{\nabla}}^2T_\text{w}$, but during the transient heating, $0>{{\nabla}}^2T_\text{c}> {{\nabla}}^2T_\text{w}$, ensuring that $\Delta T_\text{p}>0$, and this increases with time.\par
\subsection{Thermodynamics of nucleation}
\subsubsection{Energetics of cluster formation}
Due to ionic Joule heating, the liquid confined inside the nanopore enters a metastable state, and vapor cluster (or embryo) groups originate at the cylindrical pore surface and pore center [Fig.~\ref{fig:3}(a)]. These clusters are aggregates of vapor molecules that form and shrink spontaneously in the metastable liquid. Cluster formation from the bulk metastable liquid requires a minimum (reversible) work $W$. For a cluster having a critical radius $r^*$ and containing a critical number of molecules $n^*$, $W$ reaches a maximum value. Clusters having a larger number of molecules than this cluster will grow spontaneously as the free-energy decreases~\cite{debenedetti1996metastable}. $W(n^*)$ decreases as the superheating is increased, making it easier for a critical cluster to appear. The reversible work needed for the formation of a spherical vapor cluster of radius $r$ in the bulk liquid is given by~\cite{debenedetti1996metastable}
\begin{eqnarray}
{W_\text{ho}(r,{P_\text{v}})}=&&\left[{4\pi r^2\gamma} -\frac{4}{3}{\pi r^3\left({P_\text{v}}-P\right)}\right.\nonumber\\
&&\left.+\frac{4}{3}\pi r^3{P_\text{v}}{\mathrm{ln}\frac{{P_\text{v}}}{{P_\text{v}}^*} }\right]
\label{eq:four}
\end{eqnarray}
in which the superscript `$*$' denotes the properties of a critical-sized cluster. In this equation, $P$ is the atmospheric pressure, $P_\text{v}$ is the vapor pressure inside the bubble, and $\gamma$ is the surface tension at the cluster interface, which is evaluated at the cluster temperature, $T_\text{v}$.\par
The vapor pressure of the critical cluster, ${P_\text{v}}^*$, is related to the saturation vapor pressure, $P_\text{sat}$ and the critical cluster temperature, ${T_\text{v}}^*$ according to ${P_\text{v}}^* = P_\text{sat}\mathrm{exp}\left[{(P-P_\text{sat})M_\text{w}}/{(N_\text{A}\rho k{T_\text{v}}^*})\right]$. Here, $P_\text{sat}$ and the liquid density $\rho$ are evaluated at ${T_\text{v}}^*$ assuming saturation conditions. $M_\text{w}$, $N_\text{A}$, and $k$ denote the molecular weight of water, Avogadro's number, and Boltzmann's constant, respectively.
Assuming the critical cluster to be in thermal equilibrium with the liquid surrounding it~\cite{wu2010thermodynamic}, ${T_\text{v}}^*$ is calculated according to
\begin{equation}
\kappa \iint\limits_{S_\text{b}}\frac{\left(T-{T_\text{v}}^*\right)}{\delta_{\delta}}dS = 0,
\label{eq:five}
\end{equation}
where $T$ is the liquid temperature inside the pore obtained from the Joule heating simulations and interpolated over the liquid--vapor interface of the bubble, $S_\text{b}$. We assume a linear temperature gradient across a uniform thermal boundary-layer thickness, ${\delta_{\delta}}$ at $S_\text{b}$.
\begin{figure}[b]
	\includegraphics[height=4 in,width=2.3 in,angle=0]{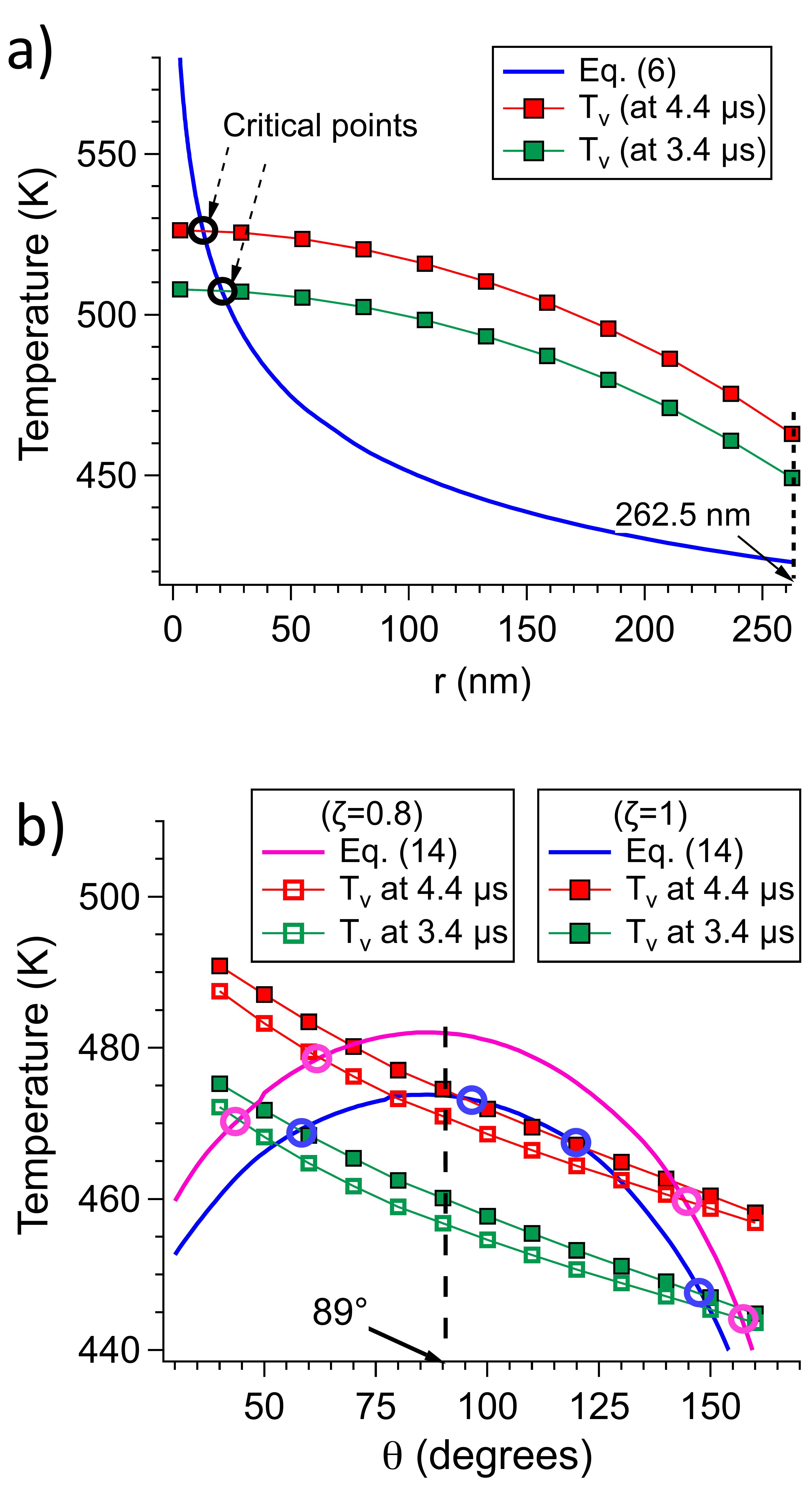}
	\caption{\label{fig:37} Variation of cluster temperature with size according to Laplace equilibrium (solid line) and thermal equilibrium (line with markers) for (a) homogeneous bubbles having radius $r$, and (b) heterogeneous bubbles having $\zeta = 1$ and $\zeta = 0.8$ inside a $D_\text{p} = 525$~nm pore. The solid and hollow square markers show the variations of vapor temperature inside bubbles having $\zeta = 1$ and $\zeta = 0.8$, respectively, when they are in thermal equilibrium with the superheated liquid inside the pore for specific time points [Eq.~(\ref{eq:five})]. The liquid temperature distribution was evaluated through Joule-heating simulations for $V_\text{app} = 7.08$~V across a $D_\text{p} = 525$~nm pore. The round hollow markers indicate the points where both Laplace and thermal equilibrium are satisfied, indicating the size and temperature inside a critical nucleus.}

\end{figure}
 Assuming the critical cluster to be in mechanical equilibrium, ${P_\text{v}}^*$ is related to the cluster radius through the Young--Laplace equation
\begin{equation}
{{P_\text{v}}^* = P + {\gamma}^*K},
\label{eq:six}
\end{equation}
where $K$ is the curvature of the bubble surface, which for a spherical homogeneous bubble is given by $K = 2/r^*$. As ${P_\text{v}}^*$ and ${\gamma}^*$ are both temperature-dependent properties, we can express the critical radius $r^*$ as a function of its temperature, which is shown by the blue curve in Fig.~\ref{fig:37}(a). Conversely, through Eq.~(\ref{eq:five}) we can express the vapor temperature of clusters of various sizes located at the pore center given the liquid temperature distribution at any given time [Fig.~\ref{fig:37}(a)]. The critical points [marked by circles in Fig.~\ref{fig:37}(a)] where these two curves intersect give us the solutions for the radius and temperature of critical clusters where both Laplace and thermal equilibrium are satisfied.
In our model, we assume that all cluster sizes have the same pressure and temperature as a critical cluster, i.e., ${P_\text{v}} = {P_\text{v}}^*$ and ${T_\text{v}} = {T_\text{v}}^*$. This means that they have the same chemical potential. The free-energy of a homogeneous cluster can be expressed as
\begin{equation}
{W_\text{ho}(n)=4\pi{\gamma}^*\left(\frac{3k{T_\text{v}}^*n}{4\pi {P_\text{v}}^*} \right)^\frac{2}{3}-nk{T_\text{v}}^*\left(1-\frac{P}{{P_\text{v}}^*}\right)},
\label{eq:seven}
\end{equation}
where $n$ is the number of molecules inside the cluster, and according to ideal gas law, this is given by
\begin{equation}
n=\frac{{P_\text{v}}V_\text{b}}{k{T_\text{v}}}\approx\frac{{P_\text{v}}^*V_\text{b}}{k{T_\text{v}}^*},
\label{eq:eight}
\end{equation}
where $V_\text{b}$ is given by
\begin{eqnarray}
{V_\text{b}}=
\begin{cases}
{\frac{4}{3}\pi r^3}, & \text{for }
\!\begin{aligned}[t]
\text{a homogeneous cluster}
\end{aligned}
\\
\frac{\pi}{3} r^3f(\theta), & \text{for a heterogeneous cluster}
\end{cases}
\label{eq:nine}
\end{eqnarray}
where $f(\theta)=2+3\cos(\theta)-\cos^{3}(\theta)$ and $\theta$ represents the contact angle formed by the heterogeneous bubble at the vapor--solid interface on the liquid side, as shown in Fig.~\ref{fig:1}(d). For heterogeneous bubble nucleation on cylindrical pore surface, we can express the reversible work for cluster formation as
\begin{multline}
W_\text{he}(r,\theta)={\gamma}^*\pi r^2\left[f(\theta)+\varepsilon\frac{\sin^{4}(\theta)}{4}\right] \\
-\frac{\pi}{3}r^3\left({P_\text{v}}^*-P\right)f(\theta).
\label{eq:ten}
\end{multline}
Here, the first term corresponds to the surface free-energy associated with the creation of the bubble and the second term is associated with the volume contribution. Due to the curvature of the cylindrical pore surface, the shape of the bubble becomes distorted from the equilibrium shape of a spherical cap on a flat surface. Here, $\varepsilon = -r/R_\text{p}$ is the perturbation parameter associated with the deformed surface. Soleimani \textit{et al.}~\cite{soleimani2013bubbles} approximately solved for the bubble/drop shape on a cylindrical surface in the absence of gravity by perturbing a spherical cap on a flat surface subject to the constraints of constant volume, uniform interface curvature, and the validity of the Young--Dupr\'e condition at the contact line. The added term, $\varepsilon\frac{\sin^{4}(\theta)}{4}$ in Eq.~(\ref{eq:ten}) captures the perturbation in the surface energy due to the curvature of the nucleating surface. This term decreases with increasing pore diameter and vanishes completely for a flat surface, in which case, $R_\text{p}\rightarrow\infty$. Although, only the first-order solution with respect to the perturbation parameter $\varepsilon$ was considered, Soleimani \textit{et al.}~\cite{soleimani2013bubbles} were able to show a good match of interface surface area and curvature when compared with numerical solutions, even when $\mid\varepsilon\mid\to1$ for large values of $\mid\theta-90\degree\mid$. As $\theta\to90\degree$, the error increases, particularly for higher values of $\mid\varepsilon\mid$. For example, the error in the perturbation surface free-energy term was shown to increase to $\sim$21\% at $\theta = 90\degree$ when $\varepsilon = 0.5$~\cite{soleimani2013bubbles}.\par
By substituting $\varepsilon$ with $-r/R_\text{p}$ and replacing $r$ with $\left( 3n kT^*_v/\pi f(\theta)P^*_v \right)^{1/3}$ in Eq.~(\ref{eq:ten}) according to Eqs.~(\ref{eq:eight}) and (\ref{eq:nine}), we arrive at

\begin{eqnarray}
W_\text{he}(n,\theta) = &&~\pi{\gamma}^* \left(\frac{3k{T_\text{v}}^*\sqrt{f(\theta)}}{\pi {P_\text{v}}^*}\right)^\frac{2}{3}n^{\frac{2}{3}}\nonumber\\
&&-\frac{nk{T_\text{v}}^*}{{P_\text{v}}^*}\left({P_\text{v}}^*-P+\frac{3{\gamma}^* \sin^4(\theta)}{4R_\text{p}f(\theta)}\right).
\label{eq:eleven}
\end{eqnarray}
Here, the vapor pressure of a critical heterogeneous cluster, ${P_\text{v}}^*$, is given by Eq.~(\ref{eq:six}), where the mean curvature $K$ is given by~\cite{soleimani2013bubbles}
\begin{equation}
K = \frac{\left\{2+[3\varepsilon\sin^4(\theta)]/[4f(\theta)]\right\}}{r^*}.
\label{eq:twelve}
\end{equation}
In this equation, ${P_\text{v}}^*$ increases with the vapor temperature ${T_\text{v}}^*$, while ${\gamma}^*$ deceases with ${T_\text{v}}^*$. Interestingly, when we try to find the critical point by taking $\partial W_\text{he}/\partial r = 0$ in Eq.~(\ref{eq:ten}), we arrive at the same Laplace condition given by Eqs.~(\ref{eq:six}) and (\ref{eq:twelve}). Similar to the perturbation surface free-energy, the error in $K$ also increases as $\theta\to90\degree$, particularly for high values of $\mid\varepsilon\mid$. For $\varepsilon = -0.9$, an error of $\sim$18\% in $K$ was found with respect to the numerical simulations for $\theta = 90\degree$.\par
For higher values of $K$, the required nucleation temperature ${T_\text{v}}^*$ would be higher [Eq.~(\ref{eq:six})]. Also, as $\varepsilon < 0$ for a concave surface, the bubble curvature is lower than that for a flat surface for the same contact angle.\par
\begin{figure}[h]
	\includegraphics[height=2.62 in,width=3.4 in,angle=0]{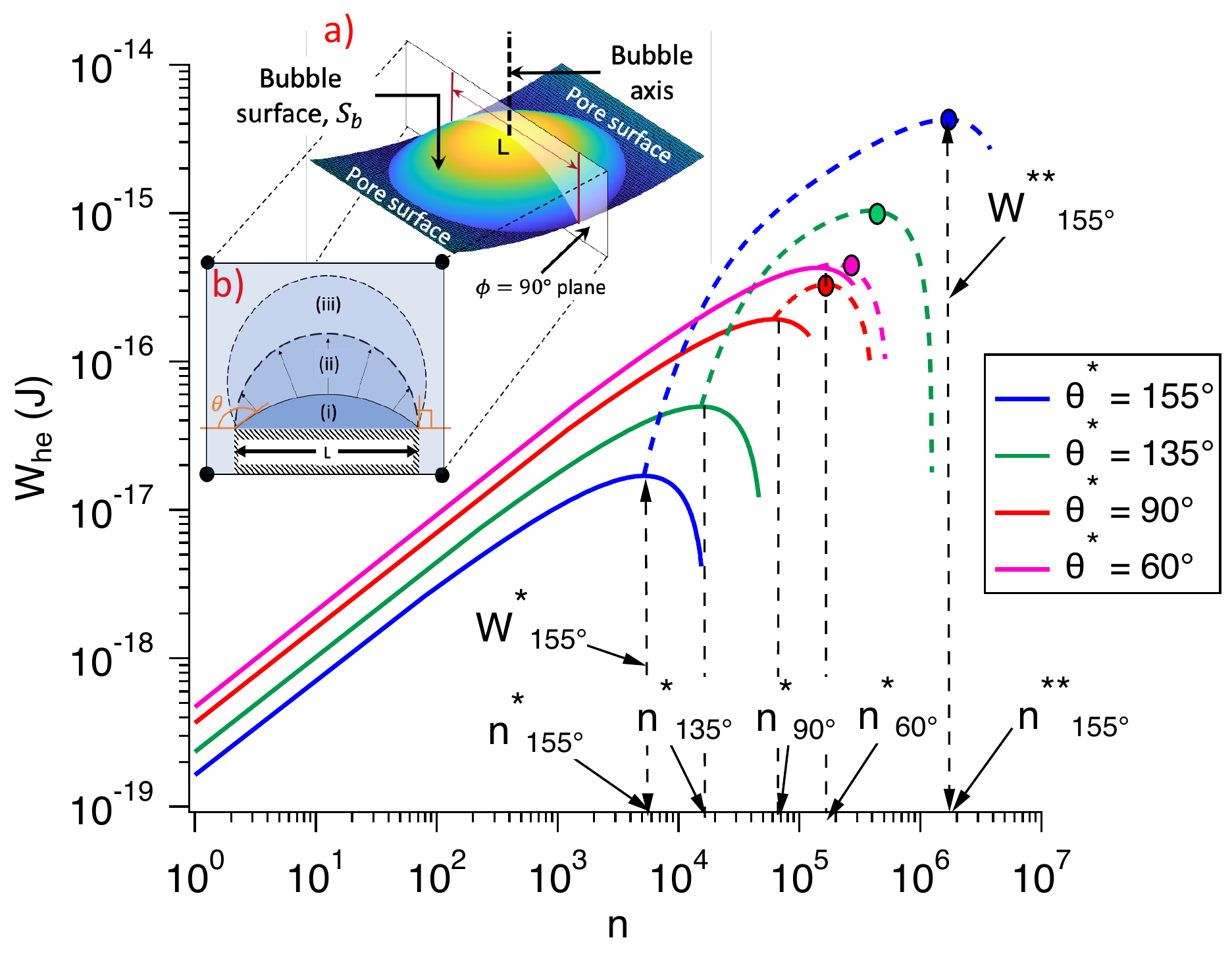}
	\caption{\label{fig:36} Solid and dotted curves showing the computed reversible work for heterogeneous cluster formation during unpinned and pinned bubble growth according to Eqs.~(\ref{eq:eleven}) and (\ref{eq:fourteen}) for $\zeta=1$ on $D_\text{p}=525$~nm pore. Inset (a) shows the 3D shape of a $\zeta = 1$ heterogeneous bubble having $\theta = 135\degree$ on a $D_\text{p}=525$~nm pore. Inset (b) shows the pinned growth of a surface nanobubble after $n = n^*$.}
\end{figure}

Unlike homogeneous nucleation, heterogeneous nucleation is constrained by the available nanostructure area~\cite{Witharana2012}. So, for a given contact angle, let the liquid--vapor interface of the bubble be given by $z_1 = z_1(a,\phi)$. Here, the bubble surface is expressed in axisymmetric co-ordinates about the bubble axis [shown in Fig.~\ref{fig:36}, inset (a)]. The parameter $a$ is a radial co-ordinate and $z_1$ is the axial distance from the bubble's center of curvature while $\phi$ is the azimuth angle. Both $a$ and $z_1$ are non-dimensionalized with respect to the bubble radius $r$. More details about the bubble shape are provided in the Supplemental Material, Sec.~5~\cite{supp}. Now, for the $\phi = 90\degree$ plane, which cuts the pore surface in a straight line, the radial distance of the intersection point of the bubble and the pore surface $a_i$ is given by imposing the condition $z_1(a_i,\phi = 90\degree) = \cos(\theta)$. Then, $a_i$ becomes a function of $\theta$, which for a flat surface is given by $a_i = \sin(\theta)$.\par
For a given $\theta$, $2a_ir^* = L$ is now the maximum coverage on the cylindrical pore surface along the $\phi = 90\degree$ plane that a heterogeneous bubble can have. In other words, the geometry of the pore imparts a constraint on the bubble size
\begin{equation}
\zeta(r,\theta) = \frac{2a_i(\theta)r}{L} \Rightarrow {\zeta}\leq 1,
\label{eq:thirteen}
\end{equation}
where $\zeta$ represents the fraction of the pore length covered by the heterogeneous bubble. As the heterogeneous bubble must fit on the cylindrical pore surface, $\zeta$ should always be less than or equal to 1.\par

Subject to mechanical equilibrium according to the Laplace equation, the nucleation temperature as a function of contact angle can be deduced as
\begin{equation}
{P_\text{v}}^* = P + {\gamma}^*\left(\frac{4a_i(\theta)}{\zeta L}-\frac{3\sin^4(\theta)}{4R_\text{p}f(\theta)}\right).
\label{eq:fourteen}
\end{equation}
We obtain the size and temperature of the critical nucleus [marked by hollow circles in Fig.~\ref{fig:37}(b)] by solving for the roots of Eqs.~(\ref{eq:fourteen}) and~(\ref{eq:five}). The Laplace equilibrium curve is shown by the blue and pink lines for $\zeta = 1$ and $\zeta = 0.8$, respectively. We find that with decreasing $\zeta$, the Laplace equilibrium curve shifts upwards, while the thermal equilibrium curve shifts downwards, causing their roots to move away from each other.\par
\begin{figure}[t]
	\includegraphics[height=3.1 in,width=3.4 in,angle=0]{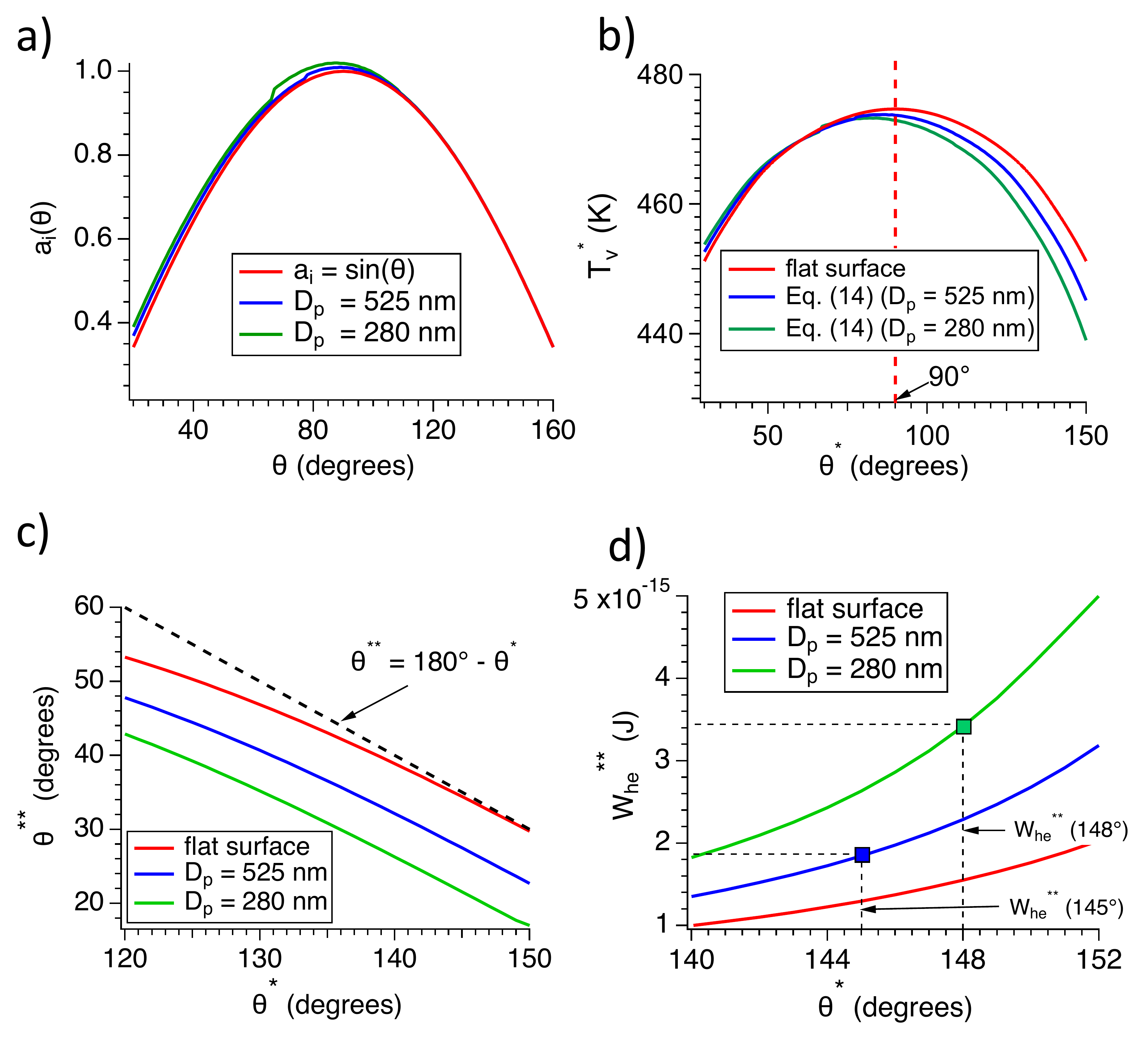}
	\caption{\label{fig:38} Variations of (a) $a_i$ and (b) vapor temperature inside a bubble with $\theta$ subject to the Laplace equation for $\zeta = 1$. Variations of (c) ${W_\text{he}}^{*}$ and (d) ${W_\text{he}}^{**}$ with ${\theta}^*$ for a $\zeta = 1$ heterogeneous bubble.}
\end{figure}

At $\zeta = 1$, for a given $\theta$, $n^*$ is the number of molecules inside a critical cluster ($r = r^*$), which is given by solving Eqs.~(\ref{eq:eight}) and (\ref{eq:fourteen}). For any value of $\theta$, when $n<n^*$, the bubble will always satisfy $\zeta < 1$. However, further addition of molecules to the critical cluster at $\zeta = 1$ results in pinned growth, during which the contact angle $\theta$ should reduce such that the constraint of $\zeta = 1$ is maintained [as shown in Fig.~\ref{fig:36}, inset (b)]. During this stage, the bubble radius varies as
\begin{equation}
r(\theta) = \frac{L}{2a_i(\theta)}.
\label{eq:sixteenpointfive}
\end{equation}
Thus, unlike the unpinned stage, during the pinned stage, $r$ and $\theta$ are not independent and are related through Eq.~(\ref{eq:sixteenpointfive}). Consequently $n$ and $W_\text{he}$ are given by combining Eq.~(\ref{eq:sixteenpointfive}) with Eqs.~(\ref{eq:eight}) and (\ref{eq:ten}), respectively. Figure~\ref{fig:36} shows the variation of $W_\text{he}$ during unpinned growth from $n = 1$ to $n = n^*$ when $\theta = {\theta}^*$ by solid lines and pinned growth for $n>n^*$ during which $\theta < {\theta}^*$ by dotted lines. Here, ${\theta}^*$ denotes the contact angle of a critical cluster. During the unpinned stage, a local maximum of free-energy $W_\text{he}$ is observed at $n = n^*$. $W_\text{he} = {W_\text{he}}^*$ at $n = n^*$. The general expression for ${W_\text{he}}^*$ is given by
\begin{equation}
{W_\text{he}}^*(\zeta,\theta) = \frac{\pi{\gamma}^*L^2{\zeta}^2f(\theta)}{12{a_i}^2}.
\label{eq:seventeen}
\end{equation}
As the term $f(\theta)/{a_i}^2$ decreases monotonically with $\theta$, ${W_\text{he}}^*$ decreases with increasing ${\theta}^*$, as shown in Fig.~\ref{fig:36}. For homogeneous nucleation, the free-energy of critical cluster depends only upon the critical cluster radius, however for heterogeneous nucleation, the critical free-energy depends upon two shape parameters, $\zeta$ and $\theta$. Without contact-line pinning, ${W_\text{he}}^*$ would have been the net free-energy barrier for heterogeneous nucleation. However, due to the pinning effect, the reversible work increases beyond $n = n^*$ until $n=n^{**}$, at which point a second maximum of $W_\text{he}$ is observed. Here, the superscript `$**$' represents the second critical point, which is reached during pinned growth. $W_\text{he} = {W_\text{he}}^{**}$ and $\theta = {\theta}^{**}$ at $n = n^{**}$. For pinned heterogeneous nucleation on a flat surface, we can easily solve for ${\theta}^{**}$ by solving for $\left|{{\frac{d{W_\text{he}}}{d\theta}}}\right|_{\theta = {\theta}^{**}} = 0$, where ${W_\text{he}}$ is given by Eq.~(\ref{eq:ten}), such that
\begin{eqnarray} \pi{\gamma}^*L^2&&\left[\frac{d}{d\theta}\left(\frac{f(\theta)}{4\sin^2{\theta}}-\frac{f(\theta)\sin({\theta}^*)}{6\sin^3{\theta}}\right)\right]=0\nonumber\\
\implies &&3\sin^5({\theta}^{**})+4\sin({\theta}^{**})\cos({\theta}^{**})\nonumber\\
&&+~6\sin({\theta}^{**})\cos^2({\theta}^{**})-2\sin({\theta}^{**})\cos^4({\theta}^{**})\nonumber\\
&&=2\sin({\theta}^*)\sin^4({\theta}^{**})+4\sin({\theta}^*)\cos({\theta}^{**})\nonumber\\
&&+~6\sin({\theta}^*)\cos^2({\theta}^{**})-2\sin({\theta}^*)\cos^4({\theta}^{**}).
\label{eq:eighteen}
\end{eqnarray}
Now, for ${\theta}$ much greater than $90\degree$, ${\theta}^{**}\approx180\degree-{\theta}^{*}$ will be the solution to Eq.~(\ref{eq:eighteen}), as $\sin^5({\theta}^{*})\approx0$ and can be neglected compared to the other terms [Fig.~\ref{fig:38}(c)]. Additionally, for a flat surface, the ${T_\text{v}}^*$ versus ${\theta}^*$ curve is symmetric about $\theta=90\degree$ [Fig.~\ref{fig:38}(b)], i.e., ${T_\text{v}}^*({\theta}^*) = {T_\text{v}}({\theta}^{**})$. So, Laplace equilibrium is satisfied at both $\theta = {\theta}^{*}$ and $\theta = {\theta}^{**}$. Also, ${W_\text{he}}^{**}({\theta}^*) = {W_\text{he}}^{*}(180\degree-{\theta}^*)$. Thus, for ${\theta}_1<{\theta}_2$, where both angles are much greater than $90\degree$, ${W_\text{he}}^{**}({\theta}_2)>{W_\text{he}}^{**}({\theta}_1)$ as ${W_\text{he}}^{*}(180\degree-{\theta}_2)>{W_\text{he}}^{*}(180\degree-{\theta}_1)$, as per Eq.~(\ref{eq:seventeen}).\par
In comparison to a flat surface, for a bubble on a curved surface having ${\theta}^*$ much greater than $90^\degree$, ${T_\text{v}}^*$ is lower due to a reduction in bubble curvature following Eq.~(\ref{eq:fourteen}) [Fig.~\ref{fig:38}(b)]. As a result, the corresponding value of ${\theta}^{**}$ for which Laplace equilibrium is re-established during pinned growth would be for ${\theta}^{**}<180\degree-{\theta}^*$, as seen in Fig.~\ref{fig:38}(c). Furthermore, as ${W_\text{he}}^{*}$ increases with decreasing ${\theta}^*$, this indicates that ${W_\text{he}}^{**}$ increases with decreasing pore diameter for a given ${\theta}^*$, as shown in Fig.~\ref{fig:38}(d).\par
This analysis also indicates that the more obtuse the bubble contact angle at the critical point, the higher the net free-energy barrier ${W_\text{he}}^{**}$. This can be noticed in both Figs.~\ref{fig:36} and~\ref{fig:38}(d).
Thus, although a critical nucleus having $n^*$ molecules is in Laplace and mechanical equilibrium, it has to overcome an additional free-energy barrier, ${W_\text{he}}^{**}-{W_\text{he}}^{*}$, to reach spontaneous growth conditions. Also, for a given value of ${\theta}^*$, the additional free-energy barrier is higher for smaller pores, making heterogeneous bubble growth comparatively more difficult than for bigger pores.
This model indicates that, due to contact-line pinning, although heterogeneous nucleation is easier at higher contact angles, it also requires a higher additional free-energy barrier to reach the spontaneous growth conditions. A similar conclusion was reached in a previous study~\cite{zou2018surface}, where obtuse contact-angled nanobubbles were shown to form more easily than their acute-angled counterparts, although the former were difficult to destabilize for nucleation. It is likely that in our system, high contact-angle surface nanobubbles may nucleate on cylindrical pore surface, but they either fail to or very rarely grow to completely block the nanopore due to the pinning effect.\par
\subsubsection{Cluster-group competition}
When the liquid inside a pore is sufficiently superheated, both homogeneous and heterogeneous clusters will appear, and, depending upon their relative free-energy requirements, a competition is expected to originate between these two cluster groups.\par
In the literature, Ostwald ripening of bubbles has been studied, which relates to the evolution of bubble size distributions post-nucleation. When a distribution of bubble sizes is formed after nucleation in an infinite medium, the larger bubbles grow at the expense of the smaller bubbles so as to reduce the total surface free-energy~\cite{watanabe2014ostwald, marqusee1984theory, de1997simple, Tomo2018, tjhung2018cluster}.
In this study, we employ a similar concept to vapor clusters originating prior to the nucleation point. During the initial range of superheating, when the pore liquid is in a metastable state but far from critical conditions, a competition may occur between the cluster groups at the cylindrical pore surface and pore center. If we consider both cluster groups as a single system and assume reversible work for cluster formation (i.e., work recovered from the collapse of either cluster group would be utilized for growth of the other group), growth of the cluster group requiring a lower $W$ would be favored. To the best of our knowledge, competition between homogeneous and heterogeneous cluster groups has not previously been studied. As the temperature increases inwards from the cylindrical pore surface, the free-energy cost of homogeneous clusters can become comparable to that of heterogeneous cluster formation. This allows the possibility of such a competition. In the traditional case of heterogeneous nucleation, where the heat flows from the surface to the liquid, liquid near the surface would always have a higher temperature than the bulk and accordingly heterogeneous clusters would be associated with a lower free-energy cost.

\begin{figure}
	\includegraphics[height=3.89 in,width=3.4 in,angle=0]{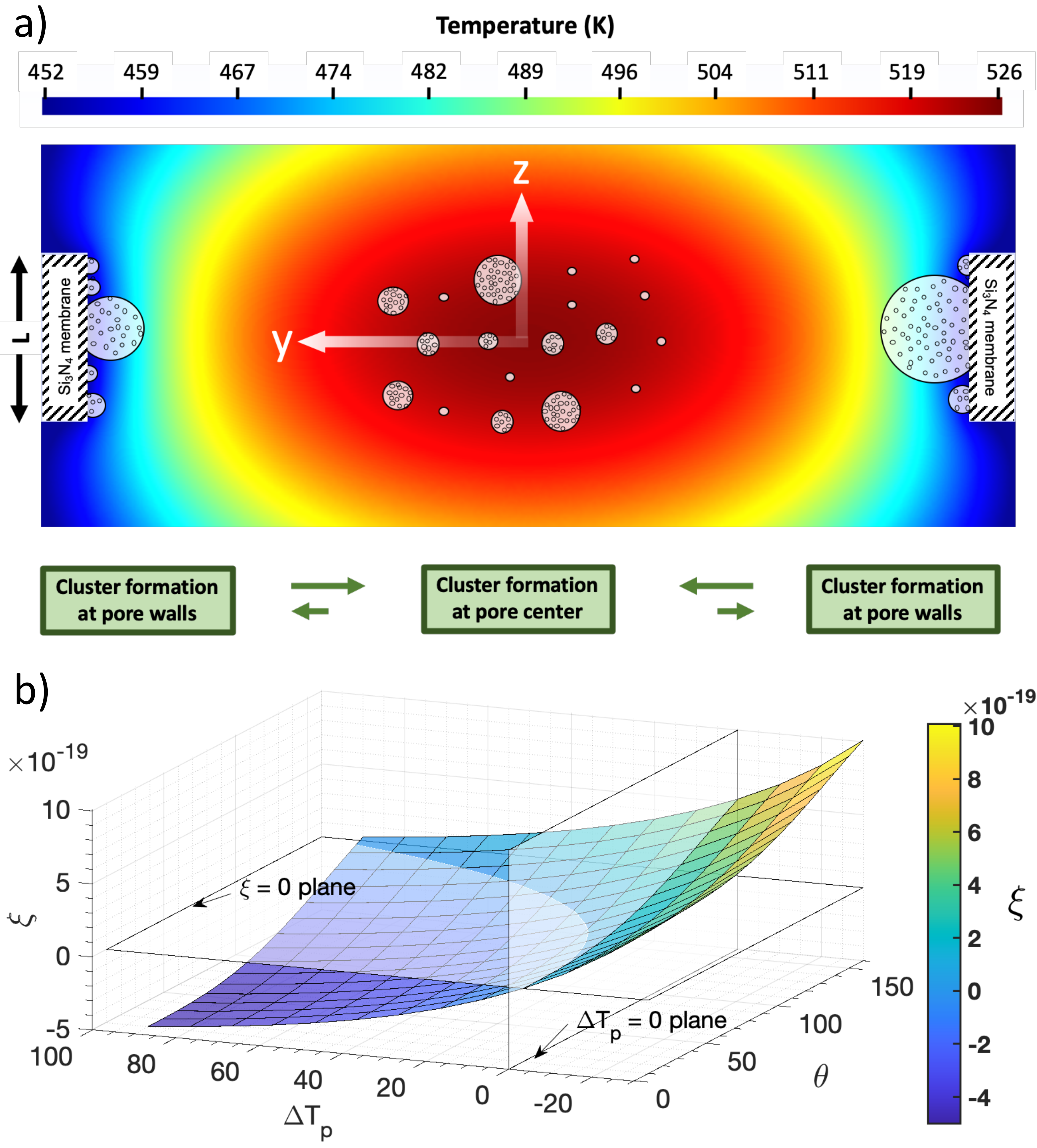}
	\caption{\label{fig:3} (a) Schematic explanation of vapor cluster formation inside a nanopore. The temperature contour corresponds to the temperature distribution at $\SI{4.4}{\micro\s}$, which was obtained through a simulation of Joule heating at 7.08~V and (b) variation of $\xi$ with $\Delta T_\text{p}$ and $\theta$ for $D_\text{p}$ = 525~nm pore. Here, $T_\text{w} = 460$~K and $T_\text{c}$ = $T_\text{w}$ + $\Delta T_\text{p}$. The homogeneous and heterogeneous clusters are at $T_\text{c}$ and $T_\text{w}$, respectively.}
\end{figure}
We hypothesize that the cluster ripening equilibrium [Fig.~\ref{fig:3}(a)] would shift towards the cluster group that would require a lower free-energy work per molecule of addition. This means that if the group of heterogeneous clusters would require a higher free-energy work to grow by ${\Delta N}$ molecules than the group of homogeneous clusters for the same number of molecules, then the homogeneous cluster group would grow at the expense of the loss of molecules from the heterogeneous clusters. In such a scenario, there is a possibility that even if the heterogeneous clusters reach conditions very close to the critical conditions, the equilibrium could shift back to homogeneous cluster ripening, thereby suppressing heterogeneous nucleation. To study this phenomenon quantitatively, we calculate $X$, which is defined as the total free-energy of either cluster group, such that
\begin{eqnarray}
{X_\text{ho}=\sum^{n^*}_{n=1}{M_\text{ho}\mathrm{\ }W_\text{ho}(n)\ \mathrm{exp}\left(-\frac{W_\text{ho}(n)}{k{T_\text{v}}^*}\right)}} \nonumber \\
{X_\text{he}(\theta)=\sum^{n^*}_{n=1}M_\text{he}(\theta)\mathrm{\ }W_\text{he}(n,\theta)} \nonumber\\
{\mathrm{\ }\mathrm{exp}\left(-\frac{W_\text{he}(n,\theta)}{k{T_\text{v}}^*}\right)},
\label{eq:nineteen}
\end{eqnarray}
Here ${T_\text{v}}^*$ for homogeneous and heterogeneous clusters are obtained from the solution of Laplace and thermal equilibrium curves [Fig.~\ref{fig:3}(a,b)]. The total number of molecules in either group is given by
\begin{eqnarray}
{N_\text{ho}={M_\text{ho}\sum^{n^*}_{n=1}\mathrm{\ }n\ \mathrm{exp}\left(-\frac{W_\text{ho}(n)}{k{T_\text{v}}^*}\right)}} \nonumber\\
{N_\text{he}(\theta)={M_\text{he}(\theta)\sum^{n^*}_{n=1}\mathrm{\ }n\mathrm{\ }\mathrm{exp}\left(-\frac{W_\text{he}(n,\theta)}{k{T_\text{v}}^*}\right)}}.
\label{eq:twenty}
\end{eqnarray}
Here, $M_\text{ho}$ and $M_\text{he}(\theta)$ are the total numbers of clusters in the homogeneous and heterogeneous cluster groups having contact angle $\theta$, respectively. We assume here that clusters belonging to either group will follow a Boltzmann distribution. As the free-energy of the clusters increases with size, the number of clusters decays exponentially. As a result, the quantity $X$ depends remarkably upon the free energies of the smaller cluster sizes belonging to each group. Now, if each cluster group is provided with $\Delta N$ additional molecules, let the change in the number of clusters of the homogeneous and heterogeneous groups be given by $M'_\text{ho}$ and $M'_\text{he}(\theta)$, respectively. It follows that
\begin{eqnarray}
\Delta N=&&{M'_\text{ho}\sum^{n^*}_{n=1}\mathrm{\ }n\mathrm{\ }\mathrm{exp}\left(-\frac{W_\text{ho}(n)}{k{T_\text{v}}^*}\right)}\nonumber\\
=&&{M'_\text{he}(\theta)\sum^{n^*}_{n=1}\mathrm{\ }n\mathrm{\ }\mathrm{exp}\left(-\frac{W_\text{he}(n,\theta)}{k{T_\text{v}}^*}\right)},
\label{eq:twentyone}
\end{eqnarray}
and the free-energy cost associated with extra cluster formation is given by
\begin{eqnarray}
\Delta X_\text{ho}={M'_\text{ho}\sum^{n^*}_{n=1}W_\text{ho}(n)\mathrm{\ }\mathrm{\ }\mathrm{exp}\left(-\frac{W_\text{ho}(n)}{k{T_\text{v}}^*}\right)}\nonumber\\
\Delta X_\text{he}(\theta)= M'_\text{he}(\theta)\sum^{n^*}_{n=1}W_\text{he}(n,\theta)\nonumber\\
\mathrm{\ }\mathrm{exp}\left(-\frac{W_\text{he}(n,\theta)}{k{T_\text{v}}^*}\right),
\label{eq:twentytwo}
\end{eqnarray}
The relative specific free-energy cost for cluster population growth can be expressed as
\begin{equation}
{\xi} = \frac{\Delta X_\text{ho}}{\Delta N}-\frac{\Delta X_\text{he}(\theta)}{\Delta N}.
\label{eq:twentythree}
\end{equation}
When $\xi<0$, the specific free-energy cost of heterogeneous cluster formation is higher than that of homogeneous cluster formation. As the equilibrium between the two cluster groups would shift in the direction having a lower specific free-energy cost, $\xi<0$ would indicate that homogeneous clusters would increase in population while heterogeneous clusters would shrink back to the liquid phase. As the total cluster population decreases, the number of critically sized clusters decreases proportionally, reducing the statistical feasibility of heterogeneous nucleation. The parameter $\xi$ essentially depends upon the vapor temperatures of the clusters in the homogeneous and heterogeneous groups and also the contact angle $\theta$ of the heterogeneous group.\par
Figure~\ref{fig:3}(b) summarizes the impact of $\theta$ and $\Delta T_\text{p}$ on $\xi$. The heterogeneous clusters originating at the cylindrical pore surface are assumed to be at $T_\text{w}$, while the homogeneous clusters originating at the pore center are assumed to be at $T_\text{c}$. We find that $\xi$ increases with decreasing $\Delta T_\text{p}$ and increasing $\theta$. With increasing temperature, the free-energy cost for cluster formation reduces, hence when $T_\text{c}$ is greater than $T_\text{w}$, it is possible that the relative free-energy cost for homogeneous cluster formation can be lower than for heterogeneous cluster formation. On the other hand, the free-energy barrier for heterogeneous nucleation decreases with $\theta$ [Eq.~(\ref{eq:seventeen})], which causes $\xi$ to increase with increasing $\theta$. When $\Delta T_\text{p} < 0$, $\xi >0$ for all $\theta$, which makes the suppression of heterogeneous nucleation impossible [Fig.~\ref{fig:3}(b)]. For $\Delta T_\text{p} > 0$, there exist $\theta$--$\Delta T_\text{p}$ ranges where $\xi <0$, thereby signifying possible suppression of heterogeneous nucleation. In the previous subsection, we showed that at a given time point, there are two solutions for the contact angle for which Laplace and thermal equilibrium are satisfied [Fig.~\ref{fig:37}(b)]. For the lower solution value of $\theta$, although $\Delta T_\text{p}$ is lower than $\Delta T_\text{p}$ for the higher solution value of $\theta$, it is still high enough to fall into the $\xi<0$ region. As a result, in most cases, it is expected that heterogeneous nucleation would be suppressed for the lower solution value of $\theta$.\par
\section{RESULTS AND DISCUSSION}
We have observed bubble nucleation and subsequent dynamics inside a nanopore by analyzing bubble-induced current-blockage signatures. Once Joule heating was started by triggering a voltage pulse, the liquid temperature inside the nanopore started rising from the ambient temperature. After an initial heating period, a bubble nucleated and grew, blocking the ion flow. During this period, the nucleated bubble first grew under the limiting effect of inertial forces due to the high superheating temperature, followed by heat-transfer-controlled growth where vapor evaporation occurred at the interface. However, due to the limited sensible heat stored in the liquid inside the nanopore, the evaporation soon ceased, and condensation began. This triggered the bubble collapse, as the vapor pressure was not sufficient to overcome the Laplace pressure. The entire duration covering bubble nucleation, growth, and collapse constituted one downward current spike. Following the bubble collapse, both ion flow and Joule heating resumed until a second bubble nucleated, repeating the cycle. Successive bubble nucleation and reheating led to transient current spikes.\par
\subsection{Early heterogeneous nucleation}
\begin{figure}[h]
	\includegraphics[height=6.31 in,width=3.4 in,angle=0]{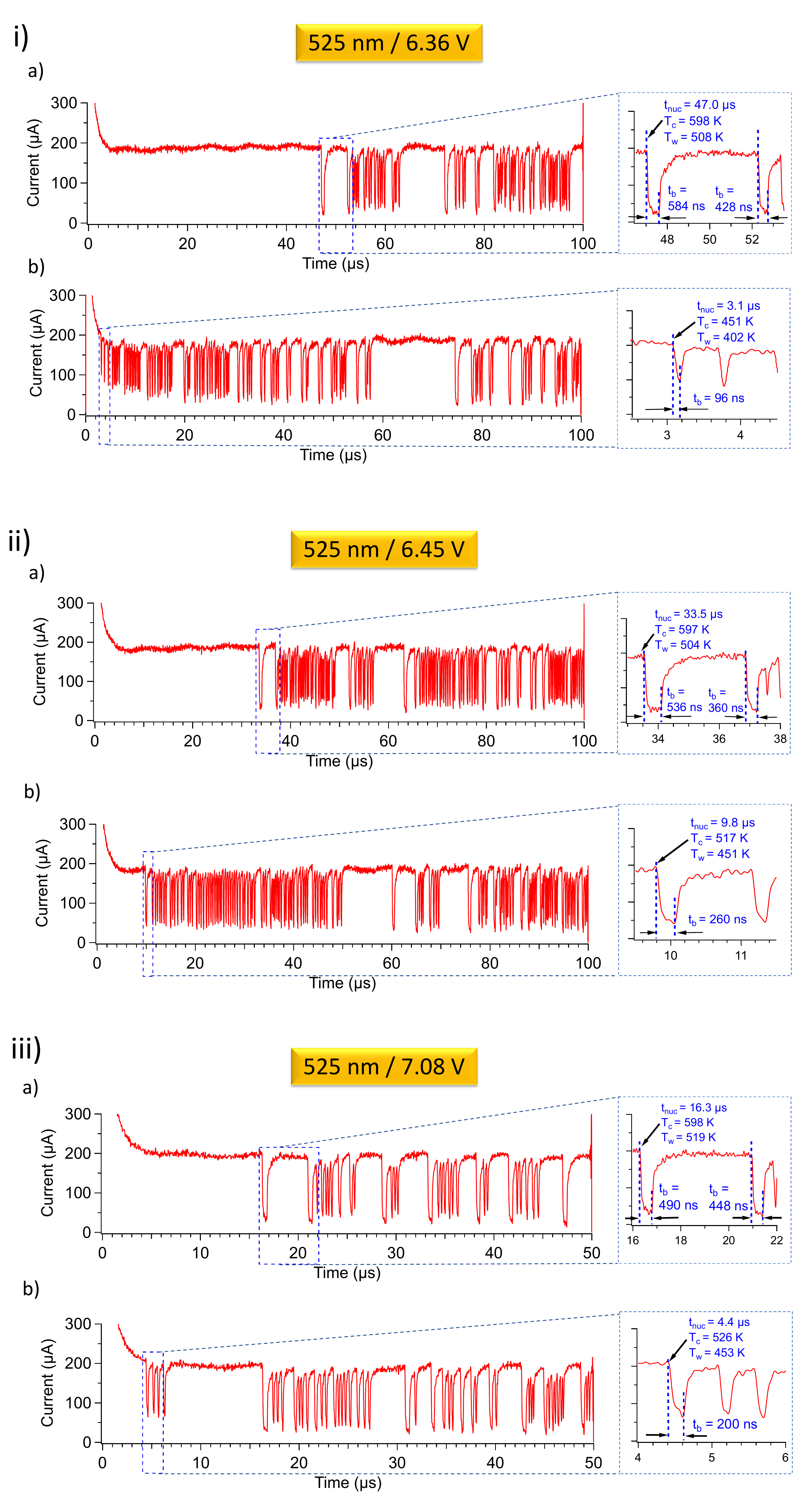}
	\caption{\label{fig:4}(i) 6.36~V, (ii) 6.45~V, and (iii) 7.08~V across a 525-nm pore. (a) Bubble signal$^\dagger$ sequence for the dominant case starting with a homogeneous nucleation followed by a mixture of homogeneous and heterogeneous bubble modes. (b) Bubble signal$^\dagger$ sequence for the rare case of early heterogeneous nucleation. ($^\dagger$Current data captured at 2.5~GS/s, sampled at 250~MHz and filtered at 12.5~MHz using an eighth-order low-pass Butterworth filter).}
\end{figure}
Figure~\ref{fig:4} shows the bubble generation characteristics of a $D_\text{p} = 525$~nm pore under bias voltages of (i) 6.36~V, (ii) 6.45~V, and (iii) 7.08~V. Note that the current signals can be nonidentical for each experimental configuration. We examined the bubble generation repeatedly (Supplemental Material, Sec.~6~\cite{supp}), and this revealed two different waiting times $t_\text{nuc}$ before the first bubble was detected. For a majority of the pulses [Case~(a)], the first bubble appeared after a long waiting time ($\SI{47}{\micro\s}$, $\SI{33.5}{\micro\s}$, and $\SI{16.3}{\micro\s}$ for 6.36~V, 6.45~V, and 7.08~V, respectively), while in rare occurrences [Case~(b)] bubble generation started much earlier ($t_\text{nuc}$ = $\SI{3.1}{\micro\s}$, $\SI{9.8}{\micro\s}$, and $\SI{4.4}{\micro\s}$ for 6.36~V, 6.45~V, and 7.08~V, respectively). Note that for the 6.45-V bias, early heterogeneous nucleation was also observed at $\SI{6.9}{\micro\s}$, as shown in the Supplemental Material, Fig.~S6.2~\cite{supp}. We estimated the transient variation of the pore-center temperature $T_\text{c}$ and the pore-wall temperature $T_\text{w}$ through Joule-heating simulations for all pore configurations under study (Supplemental Material, Fig.~S4.1~\cite{supp}). For Case~(a), the nucleation happened as long as $T_\text{c}$ reached 597~K--598~K. As the kinetic limit for homogeneous nucleation ($\sim$575~K) was exceeded, it can be concluded that the bubble was formed from the pore center. On the other hand, for Case~(b), $T_\text{c}$ was 451~K, 517~K, and 526~K, respectively, for the three voltages [Fig.~\ref{fig:4}(b)], much lower than the kinetic limit of homogeneous nucleation. It is reasonable to infer that they were heterogeneous bubbles nucleating on the cylindrical pore surface.
With regard to these early heterogeneous nucleation events, two questions arise: i) what were the conditions leading to their occurrence and ii) why, in most cases, was it suspended?
\begin{figure}[t]
	\includegraphics[height=5 in,width=2.4 in,angle=0]{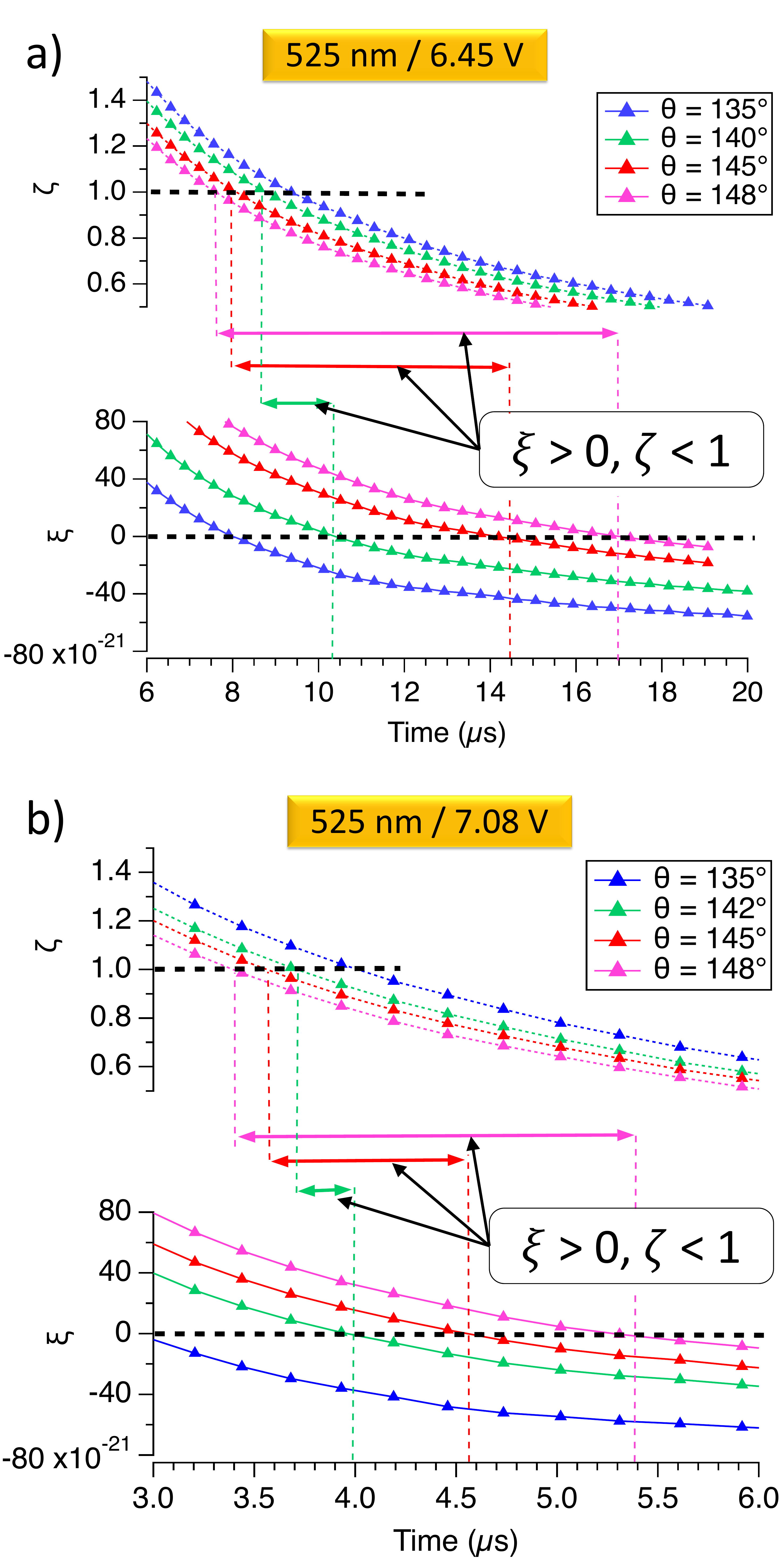}
	\caption{\label{fig:5} Transient variations of $\xi$ (solid lines) and $\zeta$ (dotted lines) for bias voltages of (a) 6.45~V and (b) 7.08~V across a 525-nm pore.}
\end{figure}
To clarify this, we calculated $\xi$ [Eq.~(\ref{eq:twentythree})] and $\zeta$ [Eq.~(\ref{eq:thirteen})], as seen in Fig.~\ref{fig:5}. At the early stages of superheating, $\xi > 0$; however, as the Joule heating progressed, more Joule heat was liberated at the pore center, causing $\Delta T_\text{p}$ to rise. Consequently, the specific free-energy for homogeneous cluster group formation decreased at a higher rate than for the heterogeneous group. Eventually, homogeneous cluster formation becomes more favorable, i.e.,$~\xi < 0$. 
For successful heterogeneous nucleation, the necessary condition of $\zeta<1$ must be satisfied before the limit of $\xi = 0$ is reached.
As Joule heating progresses, $r^*$ is reduced for a given $\theta>90\degree$ according to Eq.~(\ref{eq:six}), and thus $\zeta$ also decreases [Eq.~(\ref{eq:thirteen})]. Also, $\zeta$ decreases with $\theta$ [Eq.~(\ref{eq:thirteen})], as $a_i$ decreases with $\theta$ when $\theta >90\degree$. Meanwhile, $\xi$ increases with $\theta$, as the minimum work for heterogeneous cluster formation decreases, as per Eq.~(\ref{eq:seventeen}) and Fig.~\ref{fig:36}. Therefore, only when $\theta$ exceeds a certain value ($\sim140\degree$--$142\degree$ in Fig.~\ref{fig:5}) does there exist a time period during which the necessary conditions for heterogeneous nucleation are satisfied (i.e., $\xi > 0$ and $\zeta <1$.)\par
We summarize these valid periods in Fig.~\ref{fig:5}, which are marked by arrows.
It is shown that the period shrinks with the decrease of $\theta$. For $\theta = 135\degree$, the two conditions are never satisfied simultaneously. Note that the early heterogeneous nucleation points, $t_\text{nuc} = \SI{9.8}{\micro\s}$ in Fig.~\ref{fig:4}(ii)(b) and $t_\text{nuc} = \SI{4.4}{\micro\s}$ in Fig.~\ref{fig:4}(iii)(b), agree with the predicted periods at $\theta$ $= 140\degree$ in Fig.~\ref{fig:5}(a) and $\theta$ $= 145\degree$ in Fig.~\ref{fig:5}(b), respectively. 
After four early heterogeneous bubbles in Fig.~\ref{fig:4}(iii)(b), nucleation ceased until a homogeneous bubble appeared at $\SI{16.3}{\micro\s}$, consistent with our analysis that the heterogeneous cluster formation is suppressed due to the decrease of $\xi$ with time. As the period for heterogeneous transition enlarges with $\theta$ [Fig.~\ref{fig:5}], the early heterogeneous nucleation should be able to last much longer as $\theta$ increases. However, spontaneous heterogeneous bubble growth at high $\theta$ is highly unfavorable because the additional free-energy barrier ${W_\text{he}}^{**}$ due to contact-line pinning increases. As a result, for $\theta > 148\degree$ nanobubbles may nucleate on the pore surface, but they do not grow and have virtually no effect on the nanopore current and thus no blockage signals are registered for them during sensing experiments. Based on previous studies, the static $\theta$ on a piranha-cleaned silicon nitride surface should be in the range $20\degree$--$60\degree$~\cite{barhoumi2017silicon}. Our results indicate that when the metastability is sufficiently high, in rare situations, a heterogeneous nucleus can form at non-equilibrium $\theta$. A similar observation was reported in a previous study~\cite{german2018critical}, where, due to high supersaturation on a geometrically constrained nanoelectrode, bubble nucleation at an out-of-equilibrium $\theta$ of $150\degree$ was recorded for a platinum surface.\par
\subsection{Heterogeneous nucleation during reheating}
\begin{figure}[h]
	\includegraphics[height=3.21 in,width=3.4 in,angle=0]{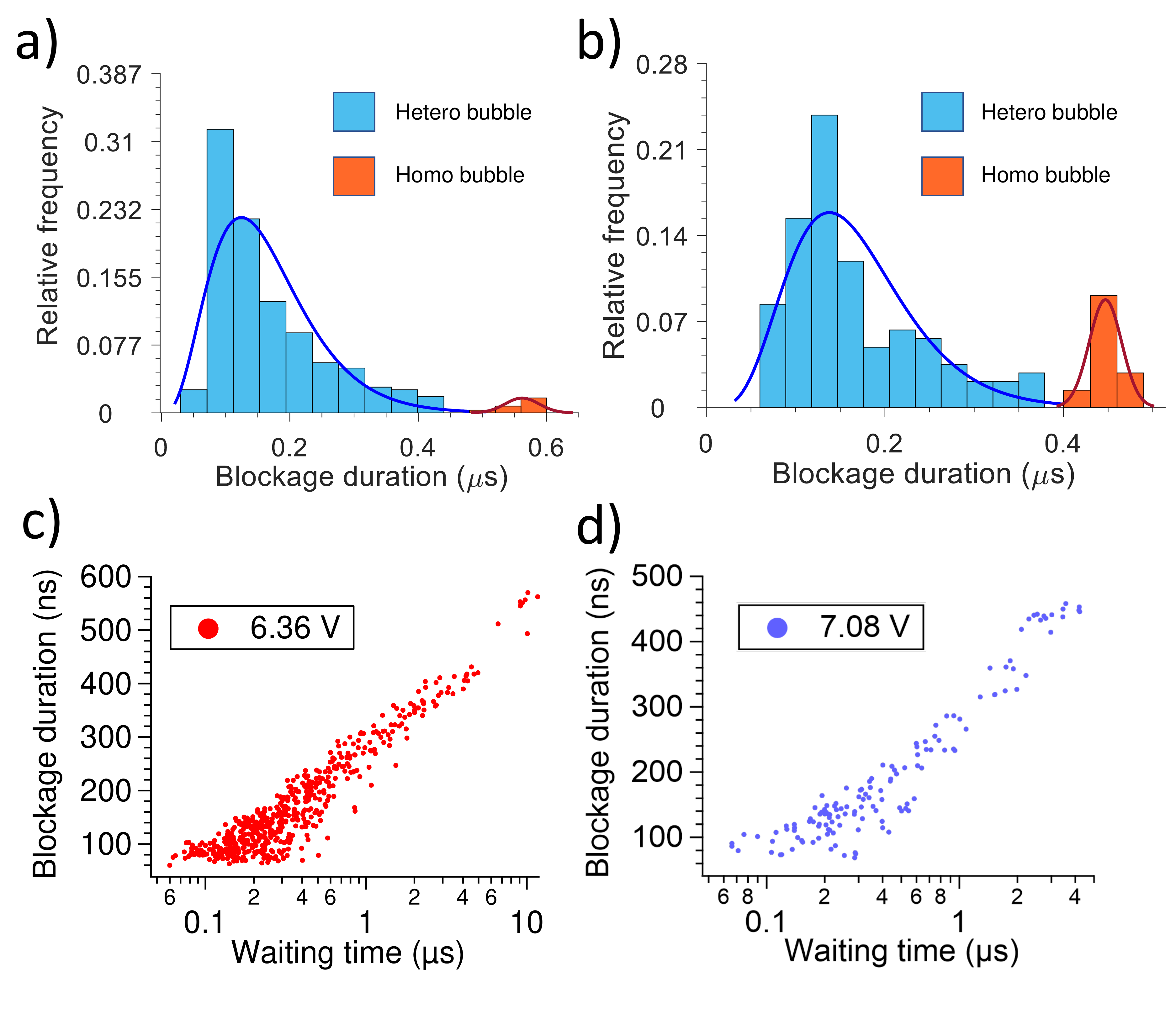}
	\caption{\label{fig:6} Histograms of bubble blockage duration for bias voltages of (a) 6.36~V and (b) 7.08~V across a 525-nm pore. Two peaks were identified, implying two modes of bubble generation. The lower peak covered high blockage-duration homogeneous bubbles (orange bars) while the higher peak covered short-duration heterogeneous bubbles (blue bars). Panels (c) and (d) show scatter plots of blockage duration versus waiting time for successive bubbles formed after the first homogeneous bubble for 6.36~V and 7.08~V, respectively.}
\end{figure}
Intriguingly, after nucleation of the first bubble, subsequent bubble events were observed, as shown in Fig.~\ref{fig:4}. However, neither the blockage duration nor the waiting times separating two bubbles were uniform (as shown in Fig.~\ref{fig:4}). To understand the root cause behind this non-periodicity, we performed statistical analysis of these signals. An in-house bubble-signal analysis package was developed in MATLAB~\cite{MATLAB:2019} to identify the blockage duration of each bubble event and the waiting time preceding it. Figures~\ref{fig:6}(a) and \ref{fig:6}(b) show histograms of the blockage duration for bubble events recorded at 6.36~V and 7.08~V for multiple pulse signals (Supplemental Material, Figs.~S6.1 and S6.3~\cite{supp}).\par
Two clear peaks are identified. The smaller peaks in Figures~\ref{fig:6}(a) and \ref{fig:6}(b) (orange bars) covers high blockage-duration bubble events while the larger peak (blue bars) demonstrates shorter blockage-duration bubbles. From the scatter plots in Figs.~\ref{fig:6}(c) and ~\ref{fig:6}(d), we find that the blockage duration of successive bubbles is proportional to the waiting time preceding its nucleation. A longer waiting time allowed more storage of sensible heat in the liquid, enabling the bubble to grow to a bigger size and thus resulting in a longer current blockage. It is also shown that the majority of bubbles had a waiting time less than $\SI{1}{\micro\s}$. As there is a strict activation temperature of $\sim$575~K for homogeneous nucleation, these early nucleation events during the reheating sequence are suspected to be heterogeneous bubbles originating from the cylindrical pore surface. Due to the lower amount of sensible heat in the liquid at the nucleation point, these heterogeneous bubbles were small and collapsed quickly (indicated by the blue bars) compared with the homogeneous bubbles (indicated by the orange bars).

The higher prevalence of heterogeneous bubbles during reheating can be traced back to a lower value of $\Delta T_\text{p}$. After the bubble collapses, the liquid inside the nanopore retains a part of the thermal energy and the post-collapse nanopore temperature is higher than the ambient temperature (Supplemental Material, Fig.~S8.1(a)~\cite{supp}). As a result, the Joule heating rate is higher during the reheating sequence and the rise of temperature back to nucleation conditions will be faster. Actually, the average post-collapse temperature within the pore volume increases with decreasing pore size. A smaller pore generates smaller bubbles, which spread out the stored sensible heat in the liquid prior to nucleation over a smaller area. As a result, the average temperature inside the pore post-collapse is usually higher for a smaller pore. According to our simulations, the average pore temperatures for the first homogeneous bubble post-collapse for the 525-nm pore under 7.08~V and the 280-nm pore under 6.84~V were 335~K and 353~K, respectively. A higher collapse temperature implies two outcomes: i) less Joule heat is needed to return to nucleation temperature, ii) the Joule heating rate is higher as electrical conductivity increases with temperature, allowing more current to flow through the nanopore, and the subsequent waiting times are reduced as a result. For the 107-nm pore used in the previous study by Golovchenko's group~\cite{Levine2016, Nagashima2014}, the waiting time observed was 117~ns, while for the 525-nm pore and the 280-nm pore, the waiting times between the first two successive homogeneous bubbles were recorded to be $\SI{4.18}{\micro\s}$ [Fig.~\ref{fig:4}(iii)(a)] and $\SI{1.36}{\micro\s}$ [Fig.~\ref{fig:8}(a)], respectively. Due to the reduction in waiting time, $\Delta T_\text{p}$ is reduced during the reheating period compared to the initial heating period for a given pore and voltage configuration according to Eq.~(\ref{eq:three}). Consequently, the probability of heterogeneous nucleation increases.\par

In most cases, when the first bubble nucleates homogeneously after a long initial waiting time, the second bubble also nucleates homogeneously after a much shorter waiting time. This trend becomes more dominant as the lifetime of the first bubble increases. For example, in Fig.~\ref{fig:4}(iii)(a), the lifetime of the first bubble is 490~ns, while the lifetime of the second bubble is 448~ns. According to the histograms in Fig.~\ref{fig:6}(b), both these bubbles fall within the homogeneous bubble blockage-duration range. As the second homogeneous bubble is short lived, the post-collapse temperature would be higher than that of the first bubble, increasing the probability of heterogeneous nucleation after the second bubble as compared to after the first bubble. Furthermore, as the heterogeneous bubbles are smaller and shorter lived, the post-collapse temperature is expected to be higher, thereby lowering the waiting time, and the accumulated $\Delta T_p$ at the next nucleation point will be lower, which makes the heterogeneous transition easier. Thus, once heterogeneous bubbles start nucleating during reheating, it becomes difficult to revert back to the homogeneous nucleation mode. As a result, the fraction of bubbles nucleating heterogeneously is considerably higher than the fraction nucleating homogeneously [Figs.~\ref{fig:6}(a) and \ref{fig:6}(b)].

Compared to the 7.08-V bias, the fraction of homogeneous bubbles is smaller for the 6.36-V bias, as a lower $H$ enables a lower value of $\Delta T_\text{p}$. As a result, heterogeneous transition becomes comparably more predominant during the reheating sequence. This phenomenon is captured in Figs.~\ref{fig:6}(a) and \ref{fig:6}(b), where the fraction of heterogeneous bubbles is larger than the fraction of homogeneous bubbles by a greater margin for the 6.36-V bias than the 7.08-V bias.\par
\begin{figure}[h]
	\includegraphics[height=0.97 in,width=3.4 in,angle=0]{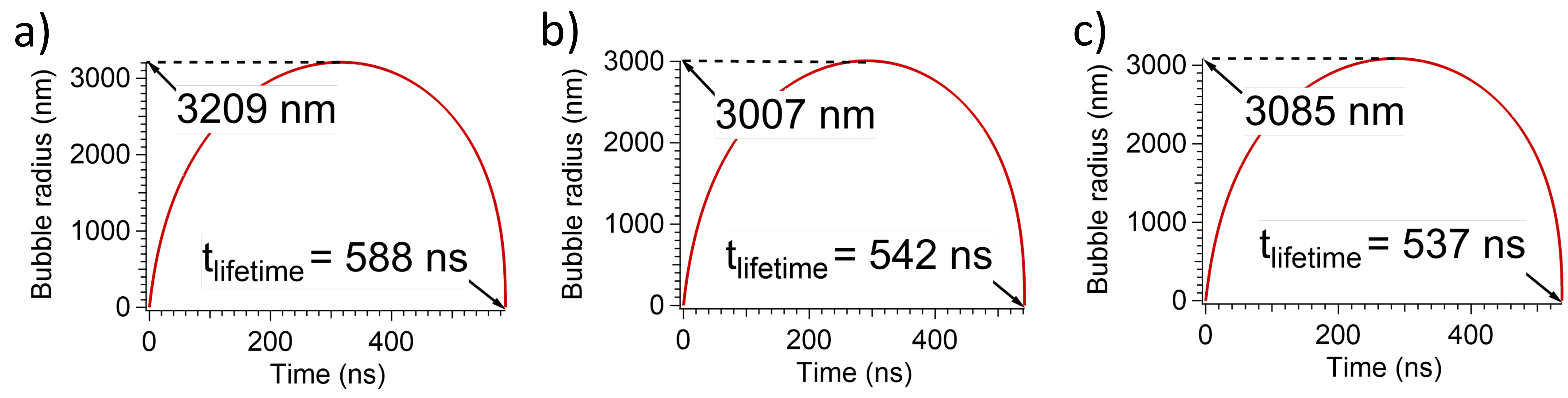}
	\caption{\label{fig:7} Panels (a), (b), and (c) show the simulated bubble growth and collapse for the first homogeneous bubble nucleating at $t_\text{nuc}$ in Fig.~\ref{fig:4}(a) for 6.36~V, 6.45~V, and 7.08~V, respectively across a 525-nm pore.}
\end{figure}
Bubble growth and collapse was simulated by solving the Rayleigh--Plesset equation over a moving-boundary one-dimensional liquid mesh~\cite{Robinson2004}. Details of the governing equations are provided in the Supplemental Material, Sec.~9~\cite{supp}. In this method, we assumed spherical bubble dynamics and the liquid temperature distribution to be symmetric along the bubble circumference. Moreover, heat dissipation through the membrane walls was neglected, which resulted in a slight overestimation of the sensible heat available in the liquid during growth. The bubble lifetime in the simulations therefore slightly overpredicts the blockage duration seen in the experiments. Figures~\ref{fig:7}(a), \ref{fig:7}(b), and \ref{fig:7}(c) show the simulated bubble growth and collapse for the first homogeneous bubble in Figs.~\ref{fig:4}(i)(a), \ref{fig:4}(ii)(a), and \ref{fig:4}(iii)(a), respectively. Compared to the experimentally observed blockage durations of 584~ns, 536~ns, and 490~ns, the simulated lifetimes were 588~ns, 542~ns, and 537~ns, respectively. Compared to the 7.08-V homogeneous bubble, the 6.36-V homogeneous bubble grows to a larger size and survives for a longer time because a lower heating rate results in a larger waiting time, thus storing more sensible heat in the liquid. This is the same reason behind the shift of the homogeneous bubble peak towards the higher blockage-duration range in the 6.36-V histogram compared to the 7.08-V histogram. The strong correlations among the waiting times, available sensible heat, and bubble lifetime are also demonstrated for heterogeneous bubbles. The early heterogeneous bubble observed at $\SI{3.1}{\micro\s}$ in Fig.~\ref{fig:4}(i)(a) (inset) has a shorter lifetime (96~ns) than the 260-ns heterogeneous bubble observed at $\SI{9.8}{\micro\s}$ in Fig.~\ref{fig:4}(ii)(a) (inset). Also, compared to this 260-ns bubble, the early heterogeneous bubble at $\SI{4.4}{\micro\s}$ in Fig.~\ref{fig:4}(iii)(a) (inset) has a shorter duration of 200~ns. In this case, although the $T_\text{w}$ is almost similar, the higher waiting time for the 6.45-V bubble allows more sensible heat to be stored in the liquid, resulting in longer-duration bubbles.\par
To validate the simulation results, we refer to the nanobubble collapse time $t_\text{c}$ model~\cite{magaletti2015shock}. The original `Rayleigh collapse' formula ~\cite{rayleigh1917viii,brennen2014cavitation}, $t_\text{c} = 0.915\left(\sfrac{\rho{R_\text{max}}^2}{[P-P_\text{v}]}\right)^\frac{1}{2}$ is valid for a macroscopic bubble and was derived through numerical integration of the Rayleigh--Plesset equation, neglecting thermal, viscous, and capillary effects. Magaletti \textit{et al.}~\cite{magaletti2015shock} modified this formula by adding the Laplace pressure term to account for the high surface tension dependence during nanobubble collapse dynamics. The modified formula is
\begin{equation}
t_\text{c} = 0.915\left(\frac{\rho{R_\text{max}}^2}{P+2\gamma/R_\text{max}-P_\text{v}}\right)^\frac{1}{2}.
\label{eq:twentyfour}
\end{equation}
The maximum bubble radius $R_\text{max}$ reached in Fig.~\ref{fig:7}(c) is 3085~nm. At this point, $\dot{R} = 0$, but the bubble interface is under a net compressive force as the vapor pressure $P_\text{v}$ is lower than the Laplace pressure, thus triggering the collapse. The vapor pressure inside the bubble and the surface tension of the interface were calculated for the interface temperature of 338~K attained at $R=R_\text{max}$, which remained almost constant for the majority of the collapse duration. According to Eq.~(\ref{eq:twentyfour}), the estimated collapse time is 256~ns, which is in good agreement with the simulated collapse time (252~ns). This validates the numerical scheme of bubble dynamics followed in this paper.

\subsection{Quasi-periodic signals for a 280-nm pore}
\begin{figure}
	\includegraphics[height=5.5 in,width=3 in,angle=0]{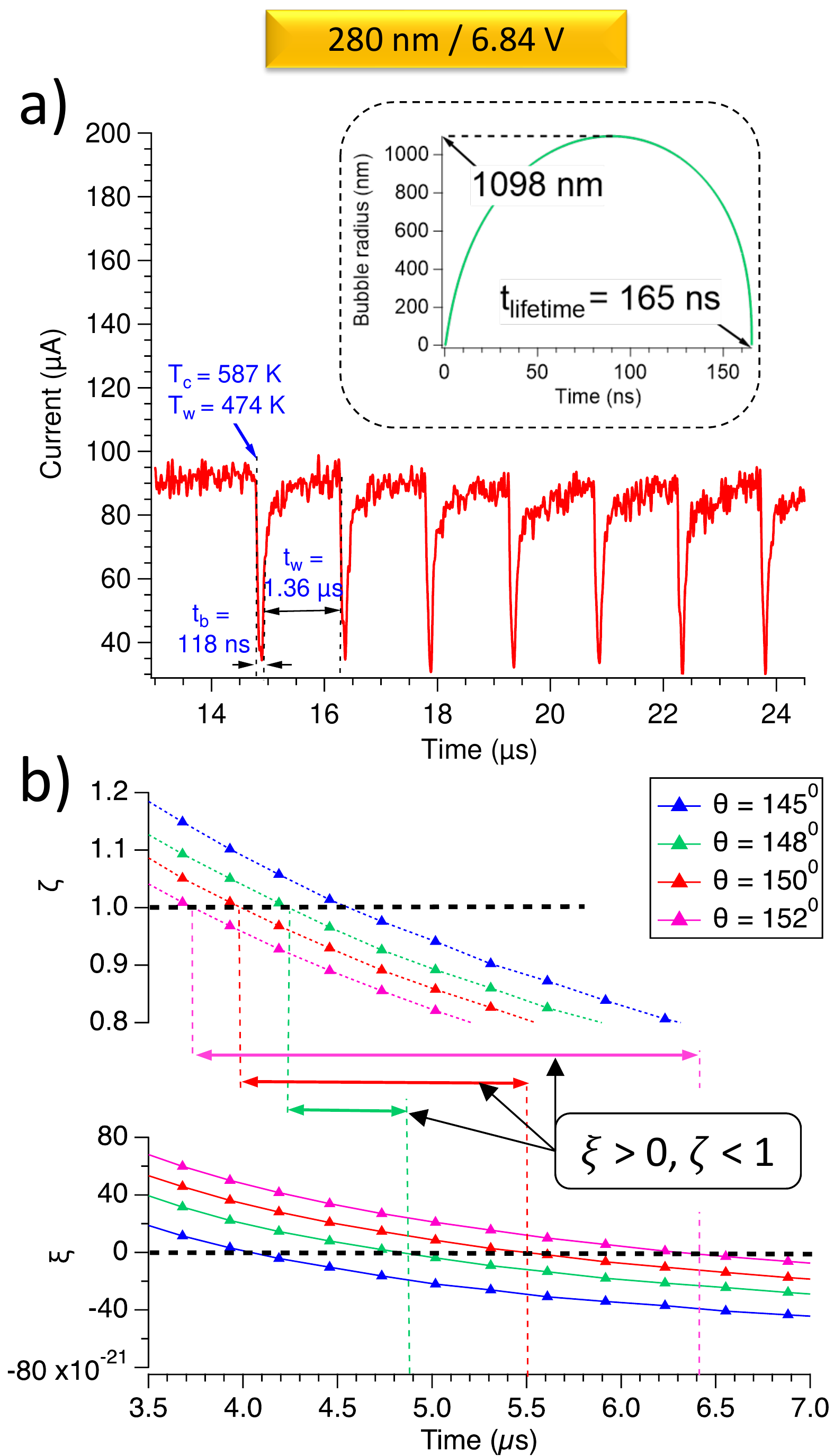}
	\caption{\label{fig:8}(a) Periodic bubble signals$^\ddagger$. Inset shows simulated bubble growth and collapse for the first homogeneous bubble nucleating at $t_\text{nuc}=14.8$ $\micro s$ and (b) transient variation of $\xi$ (solid lines) and $\zeta$ (dotted lines) for $D_p = 280$~nm under a 6.84-V bias. ($^\ddagger$Current data captured at 2.5~GS/s, sampled at 500~MHz and filtered at 25~MHz using an eighth-order low-pass Butterworth filter.)}
\end{figure}

Figure~\ref{fig:8}(a) shows the bubble signals for a 280-nm diameter pore under a bias voltage of 6.84~V. For this voltage, multi-pulse signal analysis reveals that bubble generation started consistently at $14.8 \pm \SI{0.3}{\micro\s}$ at $T_{c}=587$~K (Supplemental Material, Figs.~S6.4 and S4.1(d)~\cite{supp}). As this temperature is greater than the kinetic limit, it is indicated that the bubbles were homogeneous bubbles. The experimental blockage duration was 118~ns [Fig.~\ref{fig:8}(a)], while the simulated bubble lifetime at $t_\text{nuc}=14.8 \pm \SI{0.3}{\micro\s}$ was within $165 \pm \SI{3}{n\s}$. This overestimation may be due to the assumptions made in the model. Unlike the 525-nm pore, the rare cases of early heterogeneous nucleation were not observed for the 280-nm pore under a 6.84-V bias.
After the first bubble collapsed, the reheating continued for $\SI{1.36}{\micro\s}$ before the nucleation of the second bubble (homogeneous). The repeating of this cycle leads to periodic current-blockage signals of uniform duration and separated by consistent waiting times.\par

For the multi-pulse bubble signals (Supplemental Material, Fig.~S6.4~\cite{supp}), we found that for three out of seven pulses, only periodic homogeneous bubbles were observed, while for the remaining four pulses, transitions into non-periodic heterogeneous bubbles were observed after a few homogeneous bubbles. However, in general, periodic homogeneous bubbles were more dominant than heterogeneous bubbles in the 280-nm pore (in contrast to the 525-nm pore). 
Through $\xi$--$\zeta$ analysis, we investigated why the heterogeneous nucleation was largely suppressed in the 280-nm pore. As seen in Fig.~\ref{fig:8}(b), which shows the transient variations of $\xi$ and $\zeta$ during the initial heating period, the favorable time window for heterogeneous nucleation starts appearing for $\theta$ = $148\degree$, which is higher than $\theta$ = $140\degree$ for the 525-nm pore. This is caused by the steeper temperature variation in the smaller pore (i.e., a higher $\Delta T_\text{p}$), which lowers $\xi$. As the contact angles of the heterogeneous nuclei increase, the secondary free-energy barrier ${W_\text{he}}^{**}$ also increases. As shown in Fig.~\ref{fig:38}(d), even for the same value of $\theta$, a smaller pore is associated with a higher value of ${W_\text{he}}^{**}$. From Fig.~\ref{fig:38}(d), we find that ${W_\text{he}}^{**}$ for a $\theta = 148\degree$ critical cluster inside the 280-nm pore is nearly two times that of a $\theta=145\degree$ critical cluster inside the 525-nm pore. Due to the combined effects of pore curvature and higher $\Delta T_\text{p}$, the threshold for heterogeneous bubble formation is increased for a smaller pore, making the occurrence of heterogeneous bubbles during the experiments significantly lower. However, it is worth mentioning that even for this small pore size, oscillating heterogeneous bubbles similar to that reported in Hou \textit{et al.}~\cite{hou2015explosive} started occurring beyond a critical voltage. More investigation is needed to explore the mechanism in detail, which will be conducted in our future research.\par
\section{CONCLUSIONS}
 In summary, we showed that nanopore bubble generation is neither always homogeneous nor periodic. The temperature difference between the pore center and the pore surface, $\Delta T_\text{p}$, decides the likelihood of heterogeneous bubble nucleation on the cylindrical pore surface. We show that even if the pore surface temperature allows the formation of a heterogeneous nucleus, nucleation can be suppressed if $\Delta T_\text{p}>0$. We demonstrate that this behavior can be captured by a new thermodynamic parameter, $\xi$, which considers the relative free-energy costs of cluster groups originating at the pore center and the cylindrical pore surface. When both the necessary conditions of $\xi>0$ and $\zeta<1$ are met, heterogeneous nucleation may occur. In addition, we found that growth of pinned heterogeneous nuclei requires overcoming an additional free-energy barrier to reach spontaneous-growth conditions. This made heterogeneous bubbles more prevalent during reheating, when a lower value of $\Delta T_\text{p}$ allows low-contact-angle heterogeneous nuclei to appear on the cylindrical pore surface, and these are easy to destabilize. As growth of nucleated nanobubbles is largely fluctuation driven, it is stochastic by nature. However, homogeneous nucleation which has a higher free-energy barrier is kinetically controlled, causing deterministic nucleation temperatures and waiting times. To partially suppress heterogeneous nucleation, we engineered a pore with a high $\Delta T_\text{p}$ by reducing the pore diameter to 280~nm. Additionally, it was shown that the secondary energy barrier increases for smaller diameter pores, making growth of pinned heterogeneous surface nanobubbles more difficult. As a result, the 280-nm pore showed quasi-uniform and quasi-periodic generation of homogeneous bubbles.\par
Given that temperature hotspots within the size range of nucleating bubbles can be engineered inside nanopores through Joule heating, this platform provides a unique opportunity to study singe nanobubble dynamics. In this study, we relied on the four parameters obtained from resistive pulse sensing experiments namely the (i) baseline current, (ii) nucleation times, (iii) blockage durations and (iv) waiting times to identify the bubble generation schemes (homogeneous or heterogeneous). However, to gain an insight into other nanobubble characteristics, $e.g.$ bubble translocation \cite{gallino2018physics}, bubble oscillations \cite{hou2015explosive} and nanojets formed post bubble collapse \cite{supponenscaling_2016}, the combination of other experimental techniques such as acoustic sensing~\cite{supponen2019detailed}, light scattering~\cite{Nagashima2014, shi2016scattering} and direct visualization through 4D electron microscope~\cite{fu2017photoinduced} may prove useful in the future.
\section{ACKNOWLEDGMENTS}
This work was supported by the Japan Society for the Promotion of Science (JSPS) KAKENHI Grant Numbers 20H02081 and 20J22422. A part of this work was conducted at the Advanced Characterization Nanotechnology Platform of the University of Tokyo, supported by the Nanotechnology Platform of the Ministry of Education, Culture, Sports, Science, and Technology (MEXT), Japan, Grant Number JPMXP09A19UT0167.

\bibliography{bib2}

\end{document}